\documentclass[11pt,a4paper]{article}

\pdfoutput=1

\usepackage{jheppub}
\usepackage{ulem}

\def\beq{\begin{equation}}
\def\eeq{\end{equation}}
\def\bea{\begin{eqnarray}}
\def\eea{\end{eqnarray}}
\def\bwt{\begin{widetext}}
\def\ewt{\end{widetext}}
\def\nn{\nonumber}
\def\roughly#1{\mathrel{\raise.3ex\hbox
{$#1$\kern-.75em\lower1ex\hbox{$\sim$}}}}

\def\bs{B^0_s}
\def\bsbar{{\bar B}^0_s}

\def\order{\lower 1.8ex \hbox{\LARGE\~{}}}

\newcommand{\bctaunu}{b \to c \tau^- {\bar\nu}}

\def\bsmumu{b \to s \mu^+ \mu^-}

\def\bsll{b \to s \ell^+ \ell^-}
\def\bsnunubar{b \to s \nu {\bar\nu}}

\def \({\left(}
\def \){\right)}
\def \[{\left[}
\def \]{\right]}
\def \<{\left<}
\def \>{\right>}
\def \l|{\left|}
\def \r|{\right|}

\def \hc{{\rm h.c.}}

\def \nn{\nonumber}
\def \nl{\nn \\}

\def \hf{\frac{1}{2}}
\def \thf{\frac{3}{2}}

\def \s{\sqrt{2}}

\def \cB{{\cal B}}
\def \cL{{\cal L}}

\def \oa{\overline{a}}
\def \oc{\overline{c}}
\def \os{\overline{s}}

\def \oQ{\overline{Q}}
\def \oL{\overline{L}}

\def \otau{\overline{\tau}}

\def \al{\alpha}

\def \ga{\gamma}
\def \Ga{\Gamma}
\def \de{\delta}
\def \De{\Delta}
\def \ep{\epsilon}

\def \La{\Lambda}
\def \si{\sigma}

\def \ka{\kappa}

\def \SM{{\rm SM}}

\def \elm{{\rm em}}
\def \NP{{\rm NP}}
\def \expt{{\rm expt}}
\def \eff{{\rm eff}}
\def \elm{{\rm em}}

\def \MeV{{\rm MeV}}

\def \btos{b \to s}
\def \bbtosb{{\bar b}\to{\bar s}}

\def \al{\alpha}

\def \ga{\gamma}
\def \Ga{\Gamma}
\def \de{\delta}
\def \ka{\kappa}
\def \De{\Delta}
\def \ep{\epsilon}

\def \La{\Lambda}
\def \si{\sigma}

\def \etap{\eta^{(\prime)}}

\def \hf{\frac{1}{2}}

\def \s{\sqrt{2}}
\def \st{\sqrt{3}}

\begin{document}

\title{\boldmath Simultaneous Explanation of the $R_K$ and $R_{D^{(*)}}$ Puzzles: a Model Analysis}

\author[a,b]{Bhubanjyoti Bhattacharya,}
\author[c, d]{Alakabha Datta,}
\author[a]{Jean-Pascal Gu\'evin,}
\author[a]{\\David London,}
\author[a,e]{and Ryoutaro Watanabe}
\affiliation[a]{Physique des Particules, Universit\'e de Montr\'eal, \\
C.P. 6128, succ. centre-ville, Montr\'eal, QC, Canada H3C 3J7}
\affiliation[b]{Department of Physics and Astronomy, \\
Wayne State University, Detroit, MI 48201, USA}
\affiliation[c]{Department of Physics and Astronomy, \\
108 Lewis Hall, University of Mississippi, Oxford, MS 38677-1848, USA, }
\affiliation[d]{Department of Physics and Astronomy, \\
 2505 Correa Rd, University of Hawaii, Honolulu, Hi 96826, USA, }
 \affiliation[e]{Center for Theoretical Physics of the Universe, Institute for Basic
Science (IBS), Daejeon 305-811, Republic of Korea}
\emailAdd{bhujyo@wayne.edu}
\emailAdd{datta@phy.olemiss.edu}
\emailAdd{jean-pascal.guevin@umontreal.ca}
\emailAdd{london@lps.umontreal.ca}
\emailAdd{watanabe@lps.umontreal.ca}

\abstract{$R_K$ and $R_{D^{(*)}}$ are two $B$-decay measurements that presently
exhibit discrepancies with the SM. Recently, using an effective field theory
approach, it was demonstrated that a new-physics model can simultaneously
explain both the $R_K$ and $R_{D^{(*)}}$ puzzles. There are two UV completions
that can give rise to the effective Lagrangian: (i) $VB$: a vector boson that
transforms as an $SU(2)_L$ triplet, as in the SM, (ii) $U_1$: an $SU(2)_L$-singlet
vector leptoquark. In this paper, we examine these models individually. A key point
is that $VB$ contributes to $\bs$-$\bsbar$ mixing and $\tau \to 3\mu$,
while $U_1$ does not. We show that, when constraints from these processes
are taken into account, the $VB$ model is just barely viable. It predicts $\cB
(\tau^-\to\mu^-\mu^+\mu^-) \simeq 2.1 \times 10^{-8}$. This is measurable at Belle
II and LHCb, and therefore constitutes a smoking-gun signal of $VB$. For $U_1$,
there are several observables that may point to this model. Perhaps the most
interesting is the lepton-flavor-violating decay $\Upsilon(3S) \to \mu \tau$,
which has previously been overlooked in the literature.  $U_1$ predicts
$\cB(\Upsilon(3S) \to \mu \tau)|_{\rm max} = 8.0 \times 10^{-7}$. Thus, if a
large value of $\cB(\Upsilon(3S) \to \mu \tau)$ is observed -- and this should
be measurable at Belle II -- the $U_1$ model would be indicated.}


\arxivnumber{1609.09078}

\preprint{
{\flushright
UdeM-GPP-TH-16-252 \\
UMISS-HEP-2016-02 \\
UH-511-1265-2016 \\
CTPU-16-28\\
WSU-HEP-1605\\}}


\maketitle

\section{Introduction}
\label{Sec:Intro}

At present, there are several measurements of $B$ decays that may
indicate the presence of physics beyond the standard model (SM):
\begin{enumerate}

\item $\bsmumu$: The LHCb Collaboration has made measurements of $B
  \to K^* \mu^+\mu^-$ \cite{BK*mumuLHCb1,BK*mumuLHCb2} that deviate
  from the SM predictions \cite{BK*mumuSM}. The Belle Collaboration
  finds similar results \cite{BK*mumuBelle}. The main discrepancy is
  in the angular observable $P'_5$ \cite{P'5}. The significance of the
  discrepancy depends on the assumptions about the theoretical
  hadronic uncertainties
  \cite{BK*mumuhadunc1,BK*mumuhadunc2,BK*mumuhadunc3}. Indeed, it has
  been recently argued \cite{BK*mumuhaduncItalian} that, by including
  non-factorizable power corrections, the experimental results can be
  reproduced within the SM. However, the latest fits to the data
  \cite{BK*mumulatestfit1,BK*mumulatestfit2}, which take into account
  the hadronic uncertainties, find that a discrepancy is still
  present. It may reach the $4\sigma$ level.

The LHCb Collaboration has also measured the branching fraction and
performed an angular analysis of $B_s^0 \to \phi \mu^+ \mu^-$
\cite{BsphimumuLHCb1,BsphimumuLHCb2}. They find a $3.5\sigma$
disagreement with the predictions of the SM, which are based on
lattice QCD \cite{latticeQCD1,latticeQCD2} and QCD sum rules \cite{QCDsumrules}.

\item $R_K$:
  The LHCb Collaboration has found a hint of lepton
  non-universality.  They measured the ratio $R_K \equiv {\cal B}(B^+
  \to K^+ \mu^+ \mu^-)/{\cal B}(B^+ \to K^+ e^+ e^-)$ in the dilepton
  invariant mass-squared range 1 GeV$^2$ $\le q^2 \le 6$ GeV$^2$
  \cite{RKexpt}, and found
\bea
R_K^\expt = 0.745^{+0.090}_{-0.074}~{\rm (stat)} \pm 0.036~{\rm (syst)} ~.
\label{RKexp}
\eea
This differs from the SM prediction of $R_K^\SM = 1 \pm 0.01$
\cite{IsidoriRK} by $2.6\sigma$, and is referred to as the $R_K$
puzzle.

\item $R_{D^{(*)}}$: The charged-current decays $\bar{B} \to D^{(*)}
  \ell^{-} {\bar\nu}_\ell$ have been measured by the BaBar
  \cite{RD_BaBar}, Belle \cite{RD_Belle} and LHCb \cite{RD_LHCb}
  Collaborations. It is found that the values of the ratios
  $R_{D^{(*)}} \equiv {\cal B}(\bar{B} \to D^{(*)} \tau^{-}
  {\bar\nu}_\tau)/{\cal B}(\bar{B} \to D^{(*)} \ell^{-}
  {\bar\nu}_\ell)$ ($\ell = e,\mu$) considerably exceed their SM
  predictions. Assuming Gaussian distributions, and taking
    correlations into account, the experimental results and
    theoretical predictions can be combined to yield
    \cite{Dumont:2016xpj,Tanaka:2012nw}
\bea
R_D^{\rm ratio} \equiv \frac{R_D^\expt}{R_D^\SM} = 1.29 \pm 0.17 ~~,~~~~
R_{D^*}^{\rm ratio} \equiv \frac{R_{D^*}^\expt}{R_{D^*}^\SM} = 1.28 \pm 0.09 ~.
\label{indirect2}
\eea
The measured values of $R_D$ and $R_{D^*}$ represent deviations from
the SM of 1.7$\sigma$ and 3.1$\sigma$, respectively.  These are known
as the $R_D$ and $R_{D^*}$ puzzles.

\end{enumerate}
It must be stressed that, while the discrepancies in point 1 have some
amount of theoretical input, those in points 2 and 3 are quite
clean. As such, the $R_K$ and $R_{D^{(*)}}$ puzzles provide very
intriguing hints of new physics (NP)\footnote{Note that, while the
  $R_K$ and $R_{D^{(*)}}$ puzzles require lepton-non-universal NP,
  this is not necessarily true for point 1, see
  Ref.~\cite{Datta:2013kja} for example.}.

In Ref.~\cite{HS1}, Hiller and Schmaltz searched for a NP explanation
of the $R_K$ puzzle. They performed a model-independent analysis of
$\btos \ell^+ \ell^-$, considering NP operators of the form $({\bar s}
{\cal O} b)({\bar \ell} {\cal O}' \ell)$, where ${\cal O}$ and ${\cal
  O}'$ span all Lorentz structures. They found that the only NP
operator that can reproduce the experimental value of $R_K$ is
of $(V-A) \times (V-A)$ form: $({\bar s}_L \gamma_\mu
b_L)({\bar \ell}_L \gamma^\mu \ell_L)$. Subsequent fits
\cite{globalfits1,globalfits2,globalfits3}, which included both the $B
\to K^* \mu^+\mu^-$ and $R_K$ data, confirmed that such a NP operator
can also account for the $P'_5$ discrepancy. To be specific, $\bsmumu$
transitions are defined via the effective Hamiltonian
\bea
H_{\rm eff} &=& - \frac{\alpha G_F}{\s \pi} V_{tb} V_{ts}^*
      \sum_{a = 9,10} ( C_a O_a + C'_a O'_a ) ~, \nn\\
O_{9(10)} &=& [ {\bar s}_L \gamma_\mu b_L ] [ {\bar\mu} \gamma^\mu (\gamma_5) \mu ] ~,
\label{Heff}
\eea
where the primed operators are obtained by replacing $L$ with $R$. The
Wilson coefficients $C^{(\prime)}_a$ include both SM and NP
contributions.  In the fits it was shown that a NP contribution to
$\bsmumu$ is required; one of the possible solutions is $C_9^{NP} = -
C_{10}^{NP} < 0$, with $C_9^{NP}$ large. This corresponds to the
$({\bar s}_L \gamma_\mu b_L)({\bar \ell}_L \gamma^\mu \ell_L)$
operator of Ref.~\cite{HS1}.

In Ref.~\cite{GGL}, Glashow, Guadagnoli and Lane (GGL) stressed that
the NP responsible for lepton flavor non-universality will generally
also lead to lepton-flavor-violating (LFV) effects. To illustrate
this, they proposed the following explanation of the $R_K$ puzzle. The
NP is assumed to couple preferentially to the third generation
with $(V-A) \times (V-A)$ form, giving rise to the operator
\beq
\frac{G}{\Lambda_\NP^2} ({\bar b}'_L \gamma_\mu b'_L) ({\bar \tau}'_L \gamma^\mu \tau'_L) ~,
\label{GGLoperator}
\eeq
where $G = O(1)$, $G/\Lambda_\NP^2 \ll G_F$, and the primed fields
are the fermion eigenstates in the gauge basis. When one transforms to
the mass basis, this generates the operator $({\bar b}_L \gamma_\mu
s_L) ({\bar \mu}_L \gamma^\mu \mu_L)$ that contributes to $\bbtosb
\mu^+ \mu^-$. The contribution to $\bbtosb e^+ e^-$ is much smaller,
leading to a violation of lepton flavor universality. GGL's point was
that LFV decays, such as $B \to K \mu e$, $K\mu\tau$ and $\bs\to \mu
e$, $\mu\tau$, are also generated.

In Ref.~\cite{RKRD}, it was pointed out that, assuming the scale of NP
is much larger than the weak scale, the operator of
Eq.~(\ref{GGLoperator}) should be made invariant under the full
$SU(3)_C \times SU(2)_L \times U(1)_Y$ gauge group. There are two
possibilities:
\bea
{\cal O}_1^{NP} &=& \frac{G_1}{\Lambda_\NP^2} ({\bar Q}'_L \gamma_\mu Q'_L) ({\bar L}'_L \gamma^\mu L'_L) ~, \nn\\
{\cal O}_2^{NP} &=& \frac{G_2}{\Lambda_\NP^2} ({\bar Q}'_L \gamma_\mu \sigma^I Q'_L) ({\bar L}'_L \gamma^\mu \sigma^I L'_L) \nn\\
&=& \frac{G_2}{\Lambda_\NP^2} \left[
2 ({\bar Q}'^{i}_L \gamma_\mu Q'^{j}_L) ({\bar L}'^{j}_L \gamma^\mu L'^{i}_L)
- ({\bar Q}'_L \gamma_\mu Q'_L) ({\bar L}'_L \gamma^\mu L'_L) \right] ~,
\label{NPoperators}
\eea
where $G_1$ and $G_2$ are both $O(1)$, and the $\sigma^I$ are the
Pauli matrices. Here $Q' \equiv (t',b')^T$ and $L' \equiv
(\nu'_\tau,\tau')^T$. The key point is that ${\cal O}_2^{NP}$ contains
both neutral-current (NC) and charged-current (CC) interactions. The
NC and CC pieces can be used to respectively explain the $R_K$ and
$R_{D^{(*)}}$ puzzles\footnote{Other analyses of the $R_{D^{(*)}}$
  puzzle can be found in Refs.~\cite{RD1,RD2,RD3,RD4,RD5,RD6,RD7}.
  Distributions in $\bar{B} \to D^{(*)} \tau^{-} {\bar\nu}_\tau$
  decays can provide information on the type of new physics present
  in these decays \cite{distributionD1, distributionD2, distributionRW}. Efforts to simultaneously
  explain multiple flavor anomalies can be found in Refs.~\cite{anomalies1,anomalies2,anomalies3}.}.
  (Of course, while a common model of these anomalies is intriguing, it is
also more constraining than separate explanations of the two puzzles.)

This method was explored in greater detail in
Ref.~\cite{EffFT_3rdgen}. The starting point is the model-independent
effective Lagrangian based on Eq.~(\ref{NPoperators}):
\beq
{\cal L}_\NP = \frac{G_1}{\Lambda_\NP^2} ({\bar Q}'_L \gamma_\mu Q'_L) ({\bar L}'_L \gamma^\mu L'_L) +
\frac{G_2}{\Lambda_\NP^2} ({\bar Q}'_L \gamma_\mu \sigma^I Q'_L) ({\bar L}'_L \gamma^\mu \sigma^I L'_L) ~.
\label{Lgaugebasis}
\eeq
These operators are written in the gauge basis and involve only
third-generation fermions. In transforming from the gauge basis to the
mass basis, the left-handed down- and up-type quarks are operated upon
by the matrices $D$ and $U$, respectively, where the
Cabibbo-Kobayashi-Maskawa (CKM) matrix is $V_{CKM}=U^\dagger D$. The
leptons are different: neglecting the neutrino masses, the left-handed
charged and neutral leptons are both operated upon by the same matrix
$L$. In Ref.~\cite{EffFT_3rdgen} it is assumed that the
transformations $D$ and $L$ lead to mixing only between the
  second and third generations, so that they each depend on only one
  unknown theoretical parameter, respectively $\theta_D$ and
  $\theta_L$. In the mass basis, the above operators contribute to a
variety of $B$ decays. Ref.~\cite{EffFT_3rdgen} considers the
following processes/observables: (i) $b \to s \ell^+ \ell^-$ ($\ell =
\mu$, $e$): $B \to K^* \mu^+\mu^-$, $B_s^0 \to \phi \mu^+ \mu^-$,
$R_K$, (ii) $b \to c \tau^- {\bar\nu}_\tau$: $R_{D^{(*)}}$, (iii) $b
\to s \nu {\bar\nu}$: $B \to K^{(*)} \nu {\bar\nu}$
\cite{BKnunubarBaBar,BKnunubarBelle}. The experimental measurements
thus put constraints on the coefficients, which are all functions of
$G_1$, $G_2$, $\theta_D$ and $\theta_L$. When all constraints are
taken into account, it is found that the $R_K$ and $R_{D^{(*)}}$
puzzles can be simultaneously explained if $\theta_L$ is of the order
of $\pi/16$ and $\theta_D$ is very small (less than $V_{cb}$). With
these values for $\theta_L$ and $\theta_D$, one can make predictions
for the rates of other (LFV) processes, and this is done for $B \to
K^{(*)} \ell \ell'$ and $\bs\to \ell \ell'$ ($\ell \ell' = \tau\mu,
\tau e, \mu e$).

Finally, Ref.~\cite{EffFT_3rdgen} considers possible UV completions
that can give rise to ${\cal O}_2^{NP}$ [Eq.~(\ref{NPoperators})],
that is required to explain both $R_K$ and $R_{D^{(*)}}$. Its
coefficient ($G_2/\Lambda_\NP^2$) suggests that this operator is
generated by the tree-level exchange of a single particle. In this
case, there are only four possibilities for the underlying NP model:
(i) a vector boson ($VB$) that transforms as $({\bf 1},{\bf 3},0)$
under $SU(3)_C \times SU(2)_L \times U(1)_Y$, as in the SM, (ii) an
$SU(2)_L$-triplet scalar leptoquark ($S_3$) [$({\bf 3},{\bf
    3},-2/3)$], (iii) an $SU(2)_L$-singlet vector leptoquark ($U_1$)
[$({\bf 3},{\bf 1},4/3)$], (iv) an $SU(2)_L$-triplet vector leptoquark
($U_3$) [$({\bf 3},{\bf 3},4/3)$]. The vector boson generates only
${\cal O}_2^{NP}$, but the leptoquarks generate particular
combinations of ${\cal O}_1^{NP}$ and ${\cal O}_2^{NP}$ \cite{AGC}. It
is shown that the combination of ${\cal O}_1^{NP}$ and ${\cal
  O}_2^{NP}$ generated by the $S_3$ and $U_3$ leptoquarks cannot
simultaneously explain $R_K$ and $R_{D^{(*)}}$. The only possible UV
completions are therefore the $VB$ and $U_1$ models.

But this now raises a question. If the NP responsible for the $R_K$
and $R_{D^{(*)}}$ puzzles leads to the effective Lagrangian of
Eq.~(\ref{Lgaugebasis}), the underlying NP model is either $VB$ or
$U_1$. But which is it? Short of actually producing a $W'$/$Z'$ or a
leptoquark in an experiment, is there any way of distinguishing the
two models? At first glance, the answer is no. After all, the two
models lead to the same effective Lagrangian and so are
``equivalent.''  However, this is not really true. To see this, one
has to understand the difference between analyses based on an
effective field theory (EFT) and those based on models. In an EFT
analysis, one writes all effective operators of a given order; these
are considered as independent. The effective Lagrangian of
Eq.~(\ref{Lgaugebasis}) includes all four-fermion operators containing
two quarks and two leptons. One can also write four-quark and
four-lepton operators. But since these are uncorrelated with the
operators with two quarks and two leptons, and since only processes of
the type $q \to q' \ell \ell^{(\prime)}$ are studied, these other
operators are uninteresting.  But this does not hold in a model
analysis. Concretely, while both $VB$ and $U_1$ models lead to
operators with two quarks and two leptons, $VB$ also produces
four-quark and four-lepton operators at tree level. In particular, it
will contribute significantly to $\bs$-$\bsbar$ mixing and the
lepton-flavor-violating decay $\tau \to 3\mu$. These will lead to
additional constraints on $\theta_D$ and $\theta_L$, respectively.
Furthermore, while $VB$ contributes to $B \to K^{(*)} \nu {\bar\nu}$,
$U_1$ does not. The bottom line is that the experimental constraints
on the $VB$ model are more stringent than those on the $U_1$
model. Thus, the predictions for the rates of other processes can be
very different in the two models, and this may allow us to distinguish
them. It is this feature that is studied in the present paper.

We begin in Sec.~2 by reviewing the method of Ref.~\cite{EffFT_3rdgen}
for generating contributions to $R_K$ and $R_{D^{(*)}}$, as well as
the NP models in which this can occur. These include the vector-boson
model $VB$ and the $S_3$, $U_1$ and $U_3$ leptoquark models. The
experimental measurements that constrain these models are described in
Sec.~3. These include not only processes involving $\bsll$,
$\bsnunubar$ and $\bctaunu$, but also $\tau\to\mu\phi$, $\bs$-$\bsbar$
mixing and $\tau \to 3\mu$.  These experimental constraints are
applied to the models in Sec.~4. As in Ref.~\cite{EffFT_3rdgen}, we
find that $S_3$ and $U_3$ are excluded, leaving only $VB$ and $U_1$.
However, the constraints from $\bs$-$\bsbar$ mixing and $\tau
  \to 3\mu$ are so stringent that the $VB$ model is only barely viable.
In Sec.~5 we examine the predictions of $VB$ and $U_1$ for other
processes, to see if the two models can be distinguished. We find
that, in fact, there are a number of different ways of doing
this. Useful processes/observables include $\tau \to 3\mu$, $R_K$, and
a previously overlooked lepton-flavor-violating decay, $\Upsilon \to
\mu \tau$.  We conclude in Sec.~6.

\section{Models}

Including the generation indices $i,j,k,l$, the effective Lagrangian
of Eq.~(\ref{Lgaugebasis}) can be written as
\beq
{\cal L}_\NP =
\frac{G_1^{ijkl}}{\Lambda_\NP^2} ({\bar Q}^{(\prime) i}_L \gamma_\mu Q^{(\prime) j}_L) ({\bar L}^{(\prime) k}_L \gamma^\mu L^{(\prime) l}_L) +
\frac{G_2^{ijkl}}{\Lambda_\NP^2} ({\bar Q}^{(\prime) i}_L \gamma_\mu \sigma^I Q^{(\prime) j}_L)
         ({\bar L}^{(\prime) k}_L \gamma^\mu \sigma^I L^{(\prime) l}_L) ~.
\label{Lgaugebasis2}
\eeq
This holds in both the gauge and mass bases. The gauge eigenstates,
which involve only third-generation fermions, are indicated by primes
on the spinors; the mass eigenstates have no primes. In transforming
from the gauge basis to the mass basis, we have
\beq
u'_L = U u_L ~,~~ d'_L = D d_L ~,~~ \ell'_L = L \ell_L ~,~~ \nu'_L = L \nu_L ~,
\label{transformations}
\eeq
where $U$, $D$ and $L$ are $3\times 3$ unitary matrices and the
spinors $u^{(\prime)}$, $d^{(\prime)}$, $\ell^{(\prime)}$ and
$\nu^{(\prime)}$ include all three generations of fermions.  The fact
that the left-handed charged and neutral leptons are both operated
upon by the same matrix $L$ is a result of neglecting the neutrino
masses\footnote{If neutrino masses are not neglected, the matrices $L$
  and $N$ operate on the left-handed charged and neutral leptons,
  respectively, and the Pontecorvo-Maki-Nakagawa-Sakata (PMNS) matrix
  is $V_{PMNS} = N^\dagger L$.  However, in processes such as $b \to c
  \tau^- {\bar\nu}_\tau$ and $b \to s \nu {\bar\nu}$, the final-state
  neutrinos are not detected, and so one must sum over all
  neutrinos. In this case, since $V_{PMNS}$ is unitary
  ($V_{PMNS}^\dagger V_{PMNS} = 1$), its effect on these processes
  vanishes.}. The CKM matrix is given by $V_{CKM}=U^\dagger D$.  The
assumption of Ref.~\cite{EffFT_3rdgen} is that the transformations $D$
and $L$ involve only the second and third generations:
\beq
D =
\left(
\begin{array}{ccc}
1 & 0 & 0 \\
0 & \cos\theta_D & \sin\theta_D \\
0 & -\sin\theta_D & \cos\theta_D
\end{array}
\right)
~~,~~~~
L =
\left(
\begin{array}{ccc}
1 & 0 & 0 \\
0 & \cos\theta_L & \sin\theta_L \\
0 & -\sin\theta_L & \cos\theta_L
\end{array}
\right)
~.
\eeq
Because of these transformations, for the down-type quarks and charged
leptons, couplings involving the second generation (possibly
flavor-changing) are possible in the mass basis. (For the up-type
quarks, the first generation can also be involved.)

Specifically, in the mass basis we have
\beq
G_n^{ijkl} = g_n X^{ij} Y^{kl} ~,
\eeq
where $X$ and $Y$ include the transformations from the gauge to the
mass basis. The exact forms of these matrices depend on which
four-fermion operator is used. For the decay $\bsll$ we have
\bea
X &=& D^\dagger
\left(
\begin{array}{ccc}
0 & 0 & 0 \\
0 & 0 & 0 \\
0 & 0 & 1
\end{array}
\right)
D =
\left(
\begin{array}{ccc}
0 & 0 & 0 \\
0 & \sin^2\theta_D & -\sin\theta_D \cos\theta_D \\
0 & -\sin\theta_D \cos\theta_D & \cos^2\theta_D
\end{array}
\right) ~, \nn\\
Y &=& L^\dagger
\left(
\begin{array}{ccc}
0 & 0 & 0 \\
0 & 0 & 0 \\
0 & 0 & 1
\end{array}
\right)
L =
\left(
\begin{array}{ccc}
0 & 0 & 0 \\
0 & \sin^2\theta_L & -\sin\theta_L \cos\theta_L \\
0 & -\sin\theta_L \cos\theta_L & \cos^2\theta_L
\end{array}
\right) ~.
\label{XYdefs}
\eea
If up-type quarks are involved in a process (such as $b \to c \tau^-
{\bar\nu}$), one must include the transformation matrix $U$
[Eq.~(\ref{transformations})]. Because $V_{CKM}=U^\dagger D$, the
amplitude will involve factors of $V_{CKM}$ in addition to $X$ and
$Y$.

In terms of components, the effective Lagrangian is
\bea
{\cal L}_\NP &=& \frac{(G_1^{ijkl} + G_2^{ijkl})}{\Lambda_\NP^2} \left[
\left( \bar u_L^i \gamma_\mu u_L^j \right) \left( \bar \nu_L^k \gamma^\mu \nu_L^l \right)
+ \left( \bar d_L^i \gamma_\mu d_L^j \right) \left( \bar \ell_L^k \gamma^\mu \ell_L^l \right) \right] \nn\\
&& +~\frac{(G_1^{ijkl} - G_2^{ijkl})}{\Lambda_\NP^2} \left[
\left( \bar u_L^i \gamma_\mu u_L^j \right) \left( \bar \ell_L^k \gamma^\mu \ell_L^l \right)
+ \left( \bar d_L^i \gamma_\mu d_L^j \right)  \left( \bar \nu_L^k \gamma^\mu \nu_L^l \right) \right] \nn\\
&& +~2 \, \frac{G_2^{ijkl}}{\Lambda_\NP^2} \left[ \left( \bar u_L^i \gamma_\mu d_L^j \right)
\left( \bar \ell_L^k \gamma^\mu \nu_L^l \right) + \hc \right] ~.
\label{Eq:NPoperator}
\eea
For the processes of interest, the NP contributions are
\bea
\label{bsmumuNP}
\bsmumu &:&
\frac{(g_1 + g_2)}{\Lambda_\NP^2}
X^{23} Y^{22} \left( \bar s_L \gamma_\mu b_L \right)  \left( \bar \mu_L \gamma^\mu \mu_L \right) + \hc ~, \\
\label{bsnunuNP}
\bsnunubar &:&
\frac{(g_1 - g_2)}{\Lambda_\NP^2}
X^{23} Y^{kl} \left( \bar s_L \gamma_\mu b_L \right)  \left( \bar \nu_L^k \gamma^\mu \nu_L^l \right) + \hc
\quad {\hbox{for $k,l=2,3$}} ~, \\
\label{bctaunuNP}
b \to c \tau^- {\bar\nu} &:&
\frac{2 g_2}{\Lambda_\NP^2} \left[ (V_{CKM} X)^{23} Y^{3l} \left( \bar c_L \gamma_\mu b_L \right)
\left( \bar \tau_L \gamma^\mu \nu_L^l \right) + \hc \right]
\quad {\hbox{for $l=2,3$}} ~.
\eea
From these expressions, we see that there is no NP contribution to
$\bsnunubar$ ($\bsmumu$) if $g_1 = g_2$ ($g_1 = -g_2$).

In the above, the NP is described in effective field theory language,
as in Ref.~\cite{EffFT_3rdgen}. However, we are interested in
explicitly studying the models that can lead to this EFT. There are
two categories of NP models, those with new vector bosons, and those
that involve leptoquarks. Below we summarize the features of the
various models.

\subsection{SM-like vector bosons}
\label{VBmodel}

This model contains vector bosons ($VB$s) that transform as $({\bf
  1},{\bf 3},0)$ under $SU(3)_C \times SU(2)_L \times U(1)_Y$, as in
the SM. We refer to the $VB$s as $V = W'$, $Z'$.

In the gauge basis, the Lagrangian describing the couplings of the
$VB$s to left-handed third-generation fermions is
\bea
\De\cL^{}_{V} &=& g^{33}_{qV}\(\oQ^{\prime}_{L3}~\ga^\mu\si^I~Q^{\prime}_{L3}\)V^{I}_{\mu}
~+~ g^{33}_{\ell V}\(\oL'_{L3}~\ga^\mu\si^I~L'_{L3}\)V^{I}_{\mu}~,~~
\eea
where $\sigma^I$ ($I=1,2,3$) are the Pauli matrices. Once the heavy
$VB$ is integrated out, we obtain the following effective Lagrangian,
relevant for $\bsll$, $\bctaunu$ and $\bsnunubar$ decays:
\beq
\cL^\eff_{V} = - \frac{g^{33}_{qV}g^{33}_{\ell V}}{m^2_{V}}\(\oQ'_{L3}\ga^\mu
\si^I~Q'_{L3}\)\(\oL'_{L3}\ga^{}_\mu\si^I L'_{L3}\) ~.
\eeq
Comparing this with Eq.~(\ref{Lgaugebasis2}), we find
\beq
g_1 = 0 ~~,~~~~ g_2 = - g^{33}_{qV} g^{33}_{\ell V} ~.
\eeq
Note that $g_2$ can be either positive or negative in this model.

When one transforms to the mass basis, the $VB$s couple to other
generations. The $Z'$ contributes at tree level to $\bsmumu$ and
$\bsnunubar$; the $W'$ contributes at tree level to $\bctaunu$. These
contributions are given in Eqs.~(\ref{bsmumuNP})-(\ref{bctaunuNP}) for
the above values of $g_1$ and $g_2$.

The above processes all involve four-fermion operators that contain
two quarks and two leptons. But $VB$ exchange also produces four-quark
and four-lepton operators at tree level. In the gauge basis, the
corresponding effective Lagrangian is
\bea
{\cal L}_\NP^{4Q,4L} & = &
- \frac{(g^{33}_{qV})^2}{2m^2_{V}} \(\oQ'_{L3}\ga^\mu\si^I Q'_{L3}\)
\(\oQ'_{L3}\ga_\mu\si^I Q'_{L3}\) \nn\\
&& \hskip2truecm
-~\frac{(g^{33}_{\ell
    V})^2}{2m^2_{V}} \(\oL'_{L3}\ga^\mu\si^I L'_{L3}\)
\(\oL'_{L3}\ga_\mu\si^I L'_{L3}\) ~.
\label{4quark4lepton}
\eea
In the mass basis, these contribute to processes such as
$\bs$-$\bsbar$ mixing and $\tau \to 3\mu$, and their measurements can
be used to further constrain the $VB$ model.

There are a number of variants of the $VB$ model -- for example, see
Refs.~\cite{Crivellin:2015lwa, Isidori, dark,Chiang,Virto}. Note that
some of these models address the $\bsmumu$ anomalies with a $Z'$,
while others also try to explain the $R_{D^{(*)}}$ puzzle.  In some
models, new fermions are involved. This introduce additional
parameters, which can lead to more flexibility in predictions.

\subsection{Leptoquarks}

In Refs.~\cite{RD5, RDLQs} it was shown that six different types of
leptoquark (LQ) models can explain $R_{D^{(*)}}$. Of these, only four
lead to four-fermion operators of the desired $(V-A) \times (V-A)$
form: (i) a scalar $SU(2)_L$ singlet $S_1$, (ii) a scalar $SU(2)_L$
triplet $S_3$, (iii) a vector $SU(2)_L$ singlet $U_1$, (iv) a
vector $SU(2)_L$ triplet $U_3$. In general, tree-level LQ exchange
generates ${\cal O}_1^{NP}$ and ${\cal O}_2^{NP}$ \cite{AGC}. However,
different models will produce different combinations of the two
operators. Below, with the help of the identities in
Table~\ref{tab:1}, we determine these combinations for each of the
four LQ models. That is, we derive the relation between $g_1$ and
$g_2$, as well as the signs of these quantities.

Note that, unlike the $VB$ model, four-quark and four-lepton operators
are not produced in LQ models at tree level.

\begin{table}
\begin{center}
\begin{tabular}{|c|} \hline \hline
Fierz Transformations \\
\hline
$(\oa^{}_L b^c_L)(\oc^c_L d^{}_L) = -\hf(\oa^{}_L\ga^\mu d^{}_L)(\oc^c_L\ga^{}_\mu b^c_L)$ \\
$(\oa^{}_L\ga^\mu b^{}_L)(\oc^{}_L\ga^{}_\mu d^{}_L) = (\oa^{}_L\ga^\mu d^{}_L)(\oc^{}_L\ga^{}_\mu b^{}_L)$ \\
\hline \hline
Identities involving Pauli Matrices \\
\hline
$\si^2_{ij}\si^2_{kl} = \hf\de^{}_{il}\de^{}_{kj} - \hf\si^I_{il}\cdot(\si^I)^T_{kj}$ \\
$\si^I_{ij}\si^2_{jk}\si^2_{lm}\si^I_{mn} = \thf\de^{}_{in}\de^{}_{lk} + \hf\si^I_{in}\cdot(\si^I)^T_{lk}$ \\
$\de^{}_{ij}\de^{}_{kl} = \hf\de^{}_{il}\de^{}_{kj} + \hf\si^I_{il}\cdot\si^I_{kj}$ \\
$\si^I_{ij}\si^I_{kl} = \thf\de^{}_{il}\de^{}_{kj} - \hf\si^I_{il}\cdot\si^I_{kj}$ \\
\hline \hline
\end{tabular}
\end{center}
\caption{Fierz transformations and Pauli-matrix identities used in the analysis of LQ models.
\label{tab:1}}
\end{table}

\subsubsection{$SU(2)_L$-singlet scalar LQ ($S_1$)}

$S_1$ is a scalar LQ that is an $SU(2)_L$ triplet (it transforms
as $({\bf 3},{\bf 1},-2/3)$ under $SU(3)_C \times SU(2)_L \times
U(1)_Y$). In the gauge basis, the interaction Lagrangian for the $S_1$
LQ is given by \cite{RDLQs}
\bea
\De\cL^{}_{S^{}_1} &=& h^{33}_{S_1}\(\oQ^{\prime}_{L3}i\sigma^2 L^{\prime c}_{L3}\)
S_{1} + \hc \,,
\eea
where $\psi^c = C{\bar\psi}^T$ denotes a charge-conjugated fermion
field. When the heavy LQ is integrated out, we obtain the following
effective Lagrangian:
\bea
\cL^\eff_{S^{}_1} &=& \frac{\l|h^{33}_{S_1}\r|^2}{m^2_{S^{}_1}}\(\oQ^{\prime}_{L3i}
\si^2_{ij}L^{\prime c}_{L3j}\)\(\oL^{\prime c}_{L3k}\si^2_{kl}Q^{\prime }_{L3l}\) \\
&=& \frac{\l|h^{33}_{S_1}\r|^2}{4m^2_{S^{}_1}}\[\(\oQ^{\prime}_{L3}
\ga^\mu Q^{\prime}_{L3}\)\(\oL'_{L3}\ga^{}_\mu L'_{L3}\) - \(\oQ^{
\prime }_{L3}\ga^\mu\si^I Q^{\prime}_{L3}\)\(\oL'_{L3}\ga^{}_\mu\si
^I L'_{L3}\)\] ~. \nn
\eea
$SU(2)_L$ indices have been inserted in the first line. In the second
line, we have used relations from Table~\ref{tab:1} and then
suppressed the indices. Comparing this with Eq.~(\ref{Lgaugebasis2}),
we find
\beq
g_1 = - g_2 = \frac14 \l|h_{S_1}^{33}\r|^2 > 0 ~.
\eeq

When one transforms to the mass basis, the $S_1$ LQ couples to other
generations. However, because $g_1 = - g_2$, it does not contribute to
$\bsmumu$ [Eq.~(\ref{bsmumuNP})] and hence cannot explain $R_K$. So
this LQ model is not of interest to us.

\subsubsection{$SU(2)_L$-triplet scalar LQ ($S_3$)}

$S_3$ is a scalar LQ that is an $SU(2)_L$ triplet (it transforms as
$({\bf 3},{\bf 3},-2/3)$). In the gauge basis, its interaction
Lagrangian is given by \cite{RDLQs}
\bea
\De\cL^{}_{S^{}_3} &=& h^{33}_{S_3}\(\oQ^{\prime}_{L3}\si^I i\sigma^2
L^{\prime c}_{L3}\)S^{I}_{3} + \hc
\eea
Integrating out the heavy LQ, we obtain the following effective
Lagrangian:
\bea
\cL^\eff_{S^{}_3} &=& \frac{\l|h^{33}_{S_3}\r|^2}{m^2_{S^{}_3}}\(\oQ^{\prime}_{L3i}
\si^I_{ij}\si^2_{jk}L^{\prime c}_{L3k}\)\(\oL^{\prime c}_{L3l}\si^2_{lm}\si^I_{mn}Q^{\prime }_{L3n}\) \\
&=& \frac{\l|h^{33}_{S_3}\r|^2}{4m^2_{S^{}_3}}\[3\(\oQ^{\prime}_{L3}
\ga^\mu Q^{\prime}_{L3}\)\(\oL'_{L3}\ga^{}_\mu L'_{L3}\) + \(\oQ^{
\prime }_{L3}\ga^\mu\si^I Q^{\prime}_{L3}\)\(\oL'_{L3}\ga^{}_\mu\si
^I L'_{L3}\)\] ~. \nn
\eea
Comparing this with Eq.~(\ref{Lgaugebasis2}), we find
\beq
g_1 = 3 g_2 = \frac34 \l|h_{S_3}^{33}\r|^2 > 0 ~.
\eeq

When one transforms to the mass basis, the $S_3$ LQ couples to other
generations. The components of the $SU(2)_L$ triplet have $Q_{em} =
\frac23, -\frac13,-\frac43$. The $Q_{em} = \frac23$ LQ contributes to
$\bsll$ and $b \to c \tau^- {\bar\nu}$, while the $Q_{em} = -\frac13$
LQ contributes to $\bsnunubar$. These contributions are given in
Eqs.~(\ref{bsmumuNP})-(\ref{bctaunuNP}) for the above values of $g_1$
and $g_2$.

The $S_3$ LQ has been studied in Refs.~\cite{S3LQ1,S3LQ2}.

\subsubsection{$SU(2)_L$-singlet vector LQ ($U_1$)}

$U_1$ is a vector LQ that is an $SU(2)_L$ singlet (it transforms as
$({\bf 3},{\bf 1},4/3)$). Its interaction Lagrangian is given
in the gauge basis by \cite{RDLQs}
\bea
\De\cL^{}_{U^{}_1} &=& h_{U_1}^{33} \(\oQ^{\prime}_{L3}~\ga^\mu~L'_{L3}\)
U_{1\mu} + \hc
\eea
Integrating out the heavy LQ, and inserting $SU(2)_L$ indices, we
obtain the following effective Lagrangian:
\bea
\cL^\eff_{U^{}_1} &=& - \frac{\l|h_{U_1}^{33}\r|^2}{m^2_{U^{}_1}}\(\oQ^{\prime}_{L3i}
\ga^\mu\de^{}_{ij}~L'_{L3j}\)\(\oL'_{L3k}\ga^{}_\mu\de^{}_{kl}~Q^{\prime}_{L3l}\) \\
&=& -~\frac{\l|h_{U_1}^{33}\r|^2}{2m^2_{U^{}_1}}\[\(\oQ'_{L3}\ga^\mu
Q'_{L3}\)\(\oL'_{L3}\ga^\mu L'_{L3}\) + \(\oQ'_{L3}\ga^\mu\si^I Q'_{
L3}\)\(\oL'_{L3}\ga^{}_\mu\si^I L'_{L3}\)\] ~. \nn
\eea
Comparing this with Eq.~(\ref{Lgaugebasis2}), we find
\beq
g_1 = g_2 = - \frac12 \l|h_{U_1}^{33}\r|^2 < 0 ~.
\eeq

In the mass basis, the $U_1$ LQ couples to other generations and
contributes at tree level to $\bsmumu$ and $\bctaunu$. These
contributions are given in Eqs.~(\ref{bsmumuNP}) and (\ref{bctaunuNP})
for the above values of $g_1$ and $g_2$. However, because $g_1 = g_2$,
there is no contribution to $\bsnunubar$.

The $U_1$ LQ has been studied in Ref.~\cite{U1LQ}.

\subsubsection{$SU(2)_L$-triplet vector LQ ($U_3$)}

The $U_3$ LQ is a vector that is an $SU(2)_L$ triplet (it transforms
as $({\bf 3},{\bf 3},4/3)$). In the gauge basis, its interaction
Lagrangian is given by \cite{RDLQs}
\bea
\De\cL^{}_{U^{}_3} &=& h^{33}_{U_3}\(\oQ^{\prime}_{L3}~\ga^\mu~
\si^I L'_{L3}\)U^{I}_{3\mu} ~+~ \hc
\eea
When the heavy LQ is integrated out, the effective Lagrangian is
\bea
\cL^\eff_{U^{}_3} &=& -~\frac{\l|h^{33}_{U_3}\r|^2}{m^2_{U^{}_3}}
\(\oQ^{\prime}_{L3i}\ga^\mu\si^I_{ij}~L'_{L3j}\)\(\oL'_{L3k}\ga
^{}_\mu\si^I_{kl}~Q^{\prime}_{L3l}\) \\
&=& -~\frac{\l|h^{33}_{U_3}\r|^2}{2m^2_{U^{}_3}}\[3\(\oQ'_{L3}\ga^
\mu Q'_{L3}\)\(\oL'_{L3}\ga^\mu L'_{L3}\) - \(\oQ'_{L3}\ga^\mu\si^I
Q'_{L3}\)\(\oL'_{L3}\ga^{}_\mu\si^I L'_{L3}\)\] ~. \nn
\eea
Comparing this with Eq.~(\ref{Lgaugebasis2}), we find
\beq
g_1 = -3 g_2 = - \frac32 \l|h^{33}_{U_3}\r|^2 < 0 ~.
\eeq

In the mass basis, the $U_3$ LQ couples to other generations. The
components of the $SU(2)_L$ triplet have $Q_{em} = \frac53, \frac23,
-\frac13$. The $Q_{em} = \frac23$ LQ contributes to $\bsll$ and $b \to
c \tau^- {\bar\nu}$; the $Q_{em} = -\frac13$ LQ contributes to
$\bsnunubar$. These contributions are given in
Eqs.~(\ref{bsmumuNP})-(\ref{bctaunuNP}) for the above values of $g_1$
and $g_2$.

The $U_3$ LQ has been studied in Ref.~\cite{U3LQ, Sahoo:2016pet}.

\subsection{Summary}

We briefly recap the above results. We assume that the NP couples only
to the third generation in the gauge basis, and that it produces
four-fermion operators with a $(V-A) \times (V-A)$ structure. We find
that there are four NP models that contribute to both $R_K$ and
$R_{D^{(*)}}$. There are two operators, ${\cal O}_1^{NP}$ and ${\cal
  O}_2^{NP}$, shown in Eq.~(\ref{Lgaugebasis2}), whose coefficients
are $g_1$ and $g_2$. The four models contribute differently to ${\cal
  O}_1^{NP}$ and ${\cal O}_2^{NP}$:
\bea
VB &:& g_1 = 0 ~,~~ g_2 = - g^{33}_{qV} g^{33}_{\ell V} ~~,~~~~ {\hbox{$g_2$ can be positive or negative}} ~, \nn\\
S_3 &:& g_1 = 3 g_2 = \frac34 \l|h_{S_3}^{33}\r|^2 > 0 ~, \nn\\
U_1 &:& g_1 = g_2 = - \frac12 \l|h_{U_1}^{33}\r|^2 < 0 ~, \nn\\
U_3 &:& g_1 = -3 g_2 = - \frac32 \l|h^{33}_{U_3}\r|^2 < 0 ~.
\label{summary}
\eea
In Ref.~\cite{EffFT_3rdgen}, it is noted that $\lambda^{(3)}$ ($=g_2$)
is positive for the $S_3$ and $U_3$ models, but negative for $U_1$. This is
confirmed by the above.

\section{Constraints}
\label{constraints}

When one transforms to the mass basis, two new parameters are
introduced, $\theta_D$, $\theta_L$. The NP contributes to $\bsmumu$,
$\bsnunubar$ and $\bctaunu$. These contributions are given in
Eqs.~(\ref{bsmumuNP})-(\ref{bctaunuNP}); the coefficients are
(different) functions of $g_1$, $g_2$, $\theta_D$, $\theta_L$.
Another decay to which all four models contribute is $\tau\to\mu\phi$.
In addition, the $VB$ model contributes to other processes, such as
$\bs$-$\bsbar$ mixing and $\tau \to 3\mu$.  The experimental
measurements of, or limits on, these processes provide constraints on
the NP parameter space.

In order to compare models, we fix $\Lambda_\NP = 1$ TeV and assume a
common value for $2 g^{33}_{qV} g^{33}_{\ell V}$,
$\l|h_{S_3}^{33}\r|^2$, $\l|h_{U_1}^{33}\r|^2$ and
$\l|h^{33}_{U_3}\r|^2$.  We apply all the experimental constraints to
establish the allowed region in the ($\theta_D$, $\theta_L$) parameter
space. If there is no region in which all constraints overlap, the
model is excluded. For the models that are retained, we predict the
rates for other processes based on the allowed region in parameter
space. Since this region can be different for different models, it may
be possible to distinguish them.

\subsection{$\bsll$, $\bsnunubar$, $\bctaunu$}

The effective Hamiltonians for $\bsll$, $\bsnunubar$ and $\bctaunu$ are
\bea
\label{Eq:bsellell}
H_{\rm eff}(b \to s \ell_i\bar\ell_j) & = & - {\alpha G_F \over \sqrt 2 \pi} V_{tb} V_{ts}^*\,
 \left[ C_9^{ij}\, \left(\bar s_L \gamma^\mu b_L \right) \left( \bar\ell_i \gamma_\mu \ell_j \right) \right.  \nn\\
&& \hskip3truecm \left.
+~C_{10}^{ij}\, \left(\bar s_L \gamma^\mu b_L \right) \left( \bar\ell_i \gamma_\mu \gamma^5 \ell_j \right) \right] ~, \\
H_{\rm eff}(b \to s \nu_i\bar\nu_j) & = & - {\alpha G_F \over \sqrt 2 \pi} V_{tb} V_{ts}^*\,
C_L^{ij}\, \left(\bar s_L \gamma^\mu b_L \right) \left( \bar\nu_i \gamma_\mu (1-\gamma^5) \nu_j \right) ~, \\
H_{\rm eff}(b \to c \ell_i\bar\nu_j) & = & {4 G_F \over \sqrt 2} V_{cb} C_V^{ij}
\left( \bar c_L \gamma^\mu b_L \right) \left(\bar\ell_{iL} \gamma_\mu \nu_{jL} \right) ~,
\eea
where the Wilson coefficients include both the SM and NP
contributions: $C_X = C_X ({\rm SM}) + C_X ({\rm NP})$.  Comparing
with Eqs.~(\ref{bsmumuNP})-(\ref{bctaunuNP}) (and recalling that
${\cal L}_\NP$ and $H_{\rm eff}$ have opposite signs), we have
\bea
\label{C9NP}
C_9^{ij}({\rm NP}) = -C_{10}^{ij}({\rm NP}) &=& {\pi \over \sqrt 2 \alpha G_F V_{tb}V_{ts}^*} \,
\frac{(g_1 + g_2)}{\Lambda_\NP^2} X^{23} Y^{ij} ~, \\
C_L^{ij}({\rm NP}) &=& {\pi \over \sqrt 2 \alpha G_F V_{tb}V_{ts}^*} \, \frac{(g_1 - g_2)}{\Lambda_\NP^2} X^{23} Y^{ij} ~, \\
\label{CVNP}
C_V^{ij}({\rm NP}) &=& - { 1 \over 2 \sqrt 2 G_F V_{cb}} \,
\frac{2 g_2}{\Lambda_\NP^2} (V_{cs} X^{23} + V_{cb} X^{33}) Y^{ij} ~.
\eea
In the following subsections we examine the experimentally-preferred
values of the above quantities.

\subsubsection{$C_9^{\mu\mu}({\rm NP}) = -C_{10}^{\mu\mu}({\rm NP})$}

A global analysis of the $\bsll$ anomalies was recently performed in
Ref.~\cite{BK*mumulatestfit1}. The fit included data on $B \to K^{(*)}
\mu^+ \mu^-$, $B \to K^{(*)} e^+ e^-$, $\bs \to \phi \mu^+ \mu^-$, $B
\to X_s \mu^+ \mu^-$, $b \to s \gamma$ and $\bs \to \mu^+ \mu^-$. It
was found that there is a significant disagreement with the SM,
possibly as large as $4\sigma$, and that it can be explained if there
is NP in $b \to s \mu^+ \mu^-$. There are four possible explanations,
each having roughly equal goodness-of-fits: (i) $C_9^{\mu\mu}({\rm
  NP}) < 0$, (ii) $C_9^{\mu\mu}({\rm NP}) = - C_{10}^{\mu\mu}({\rm
  NP}) < 0$, (iii) $C_9^{\mu\mu}({\rm NP}) = - C_{9}^{\prime \mu\mu}({\rm
  NP}) < 0$, and (iv) $C_9^{\mu\mu}({\rm NP}) = - C_{10}^{\mu\mu}({\rm NP})
= - C_{9}^{\prime \mu\mu}({\rm NP}) = - C_{10}^{\prime \mu\mu}({\rm
  NP}) < 0$. Of these, it is solution (ii) that interests
us. According to the fit, the allowed $3\sigma$ range for the Wilson
coefficients is
\beq
-1.13 \le C_9^{\mu\mu}({\rm NP}) = -C_{10}^{\mu\mu}({\rm NP}) \le -0.21 ~.
\eeq
Note that the above range of the NP contribution is consistent with
the $R_K$ anomaly: the central value of $R_K^\expt$ can be explained
with $C_9^{\mu\mu}({\rm NP}) \simeq -0.55$.

\subsubsection{$C_L^{ij}({\rm NP})$}

$C_L^{ij}({\rm NP})$ can be constrained by the existing data of $\bar
B \to K \nu\bar\nu$ and $\bar B \to K^* \nu\bar\nu$ decays.  The BaBar
and Belle Collaborations give the following 90\% C.L. upper limits
\cite{BKnunubarBaBar,BKnunubarBelle}:
\bea
\mathcal{B}(B^+ \to K^+ \nu\bar\nu) & \le & 1.7 \times 10^{-5} ~, \nn\\
\mathcal{B}( B^+ \to K^{*+} \nu\bar\nu) & \le & 4.0 \times 10^{-5} ~, \nn\\
\mathcal{B}( B^0 \to K^{*0} \nu\bar\nu) & \le & 5.5 \times 10^{-5} ~.
\eea
In Ref.~\cite{Buras:2014fpa}, these are compared with the SM
predictions
\bea
\mathcal{B}_K^{\rm SM} \equiv \mathcal{B}(B \to K \nu\bar\nu)_{\rm SM} = (3.98 \pm 0.43 \pm 0.19) \times 10^{-6} ~, \nn\\
\mathcal{B}_{K^*}^{\rm SM} \equiv \mathcal{B}(B \to K^{*} \nu\bar\nu)_{\rm SM} = (9.19 \pm 0.86 \pm 0.50) \times 10^{-6} ~.
\eea
Taking into account the theoretical uncertainties
\cite{Buras:2014fpa}, the 90\% C.L. upper bounds on the NP
contributions are
\beq
\frac{\mathcal B_K^{{\rm SM} + {\rm NP}}}{\mathcal B_K^{\rm SM}} \le 4.8 ~~,~~~~
\frac{\mathcal B_{K^*}^{{\rm SM} + {\rm NP}}}{\mathcal B_{K^*}^{\rm SM}} \le 4.9 ~.
\eeq
We have
\beq
 \frac{\mathcal B_K^{{\rm SM} + {\rm NP}}}{\mathcal B_K^{\rm SM}}
 = \frac{\mathcal B_{K^*}^{{\rm SM} + {\rm NP}}}{\mathcal B_{K^*}^{\rm SM}}
 = \frac{1}{3 |C_L^{\rm SM}|^2} \left( 3 |C_L^{\rm SM}|^2 +2 C_L^{\rm SM} \sum_{i=1}^3 {\rm Re}[C_L^{ii}({\rm NP})] + \sum_{i,j=1}^3 |C_L^{ij}({\rm NP})|^2   \right) ~,
\eeq
where $C_L^{\rm SM} \simeq -1.47 / \sin^2\theta_W \simeq - 6.36$
($\theta_W$ is the Weinberg angle).  The bound on NP therefore becomes
\beq
-13 \sum_{i=1}^3 {\rm Re}[C_L^{ii}({\rm NP})] + \sum_{i,j=1}^3 |C_L^{ij}({\rm NP})|^2 \le 473 ~.
\eeq

A constraint on the NP contribution can also be obtained from the
inclusive decay. The ALEPH Collaboration gives the 90\% C.L. upper
limit as $\mathcal{B}( B \to X_s \nu\bar\nu) \le 6.4 \times 10^{-4}$
\cite{Barate:2000rc}. However, this implies $\mathcal B_{X_s}^{{\rm
    SM} + {\rm NP}}/\mathcal B_{X_s}^{\rm SM} \le 22$, which is a weaker
constraint than that from the exclusive decays.

\subsubsection{$C_V^{\ell\nu}({\rm NP})$}

The constraint on $C_V^{ij}$ can be obtained from the comparison of
the measurements of the ratios $R_{D^{(*)}} \equiv {\cal B}(\bar{B}
\to D^{(*)} \tau^{-} {\bar\nu}_\tau)/{\cal B}(\bar{B} \to D^{(*)}
\ell^{-} {\bar\nu}_\ell)$ ($\ell = e,\mu$) with their SM
expectations. This is shown in Eq.~(\ref{indirect2}), and leads to the
$3\sigma$ bounds
\beq
0.79 \le R_D^{\rm ratio} \le 1.79  ~~,~~~~
1.02 \le R_{D^*}^{\rm ratio} \le 1.53 ~,
\label{RDconstraint}
\eeq
where
\beq
R_D^{\rm ratio} = R_{D^*}^{\rm ratio}
 =  {2\( \left|1 + C_V^{\tau \nu_\tau}({\rm NP}) \right|^2 + \displaystyle \sum_{j=1,2} \left|C_V^{\tau \nu_j}({\rm NP}) \right|^2 \)
 \over 1 + \left|1+ C_V^{\mu \nu_\mu}({\rm NP}) \right|^2 + \displaystyle \sum_{j=1,3} \left| C_V^{\mu \nu_j}({\rm NP}) \right|^2} ~.
\eeq
Here we have assumed $C_V^{e\nu_j}(\NP)=0$.

\subsection{$\tau\to\mu\phi$}

The NP effective Lagrangian of Eq.~(\ref{Eq:NPoperator}) generates the
process $\tau\to\mu s\os$:
\bea
\cL_\eff &=& \frac{g_1 + g_2}{\La^2_\NP} X^{22} Y^{23}
\(\os_L \ga^\mu s_L\)\(\otau_L \ga_\mu\mu_L \) ~,
\eea
which will lead to $\tau\to\mu\phi$ and  $\tau\to\mu\etap $. Writing the hadronic currents as
\bea
\langle 0 |\os\ga^\mu s |\phi \rangle = f^{}_\phi m^{}_\phi \ep^\mu_\phi \,, \quad
\langle 0 |\os\ga^\mu s |\etap \rangle = i f^{}_{\etap} p^\mu_{\etap} \,,
\eea
the branching ratios (neglecting the mass of the muon)  are given by
\bea
\cB(\tau\to\mu\phi) &=& \dfrac{f^2_\phi m^3_\tau\tau^{}_\tau}{128\pi\La
^4_\NP}\l|\ka\r|^2\(1 - \eta^2_\phi\)^2\(1 + 2\eta^2_\phi\) ~,~~ \nn\\
\cB(\tau\to\mu\etap) &=& \dfrac{f^2_{\etap} m^3_\tau\tau^{}_\tau}{128
\pi\La^4_\NP}\l|\ka\r|^2(1 - \eta^2_{\etap}) ~,~~
\eea
where $\ka = (g^{}_1 + g^{}_2)X^{22}Y^{23}$, $\eta_\phi \equiv
m_\phi/m_\tau$ and $\eta_\etap \equiv m_\etap/m_\tau$ . Thus we obtain
the following ratio:
\bea
\dfrac{\cB(\tau\to\mu\etap)}{\cB(\tau\to\mu\phi)} &=& \dfrac{f^2_{\etap}}
{f^2_\phi}\cdot\dfrac{1 - \eta^2_{\etap}}{(1 - \eta^2_\phi)^2(1 + 2\eta^2
_\phi)} ~.~~
\eea

We may use the following expression to estimate $f^2_\phi$:
\beq
f^2_\phi = \dfrac{27m_\phi\Ga_\phi\cB(\phi\to\mu^+\mu^-)}{4\pi
\al^2_\elm} ~.
\eeq
Taking the values for $m_\phi$, $m_\tau$, $\tau_\tau$, $\Ga_\phi$ and
$\cB(\phi\to\mu^+\mu^-)$ from Ref.~\cite{pdg}, this yields $f_\phi
\approx 225~\MeV$.  For the $\etap$ decay constant we get (using
$f_\pi = 130$ MeV, $f^{} _1 \sim 1.1 f^{}_\pi$, $f^{}_8 \sim 1.3
f^{}_\pi$ \cite{fetap} , and $\theta = 19.5^ \circ$\cite{etamix})
\bea
f^{}_\eta &=& -\dfrac{f^{}_\pi}{\st}\(\s\cos\theta\dfrac{f^{}_8}{f^{}_\pi}
+ \sin\theta\dfrac{f^{}_1}{f^{}_\pi}\) \simeq - 157.63~\MeV ~,~~ \nn\\
f^{}_{\eta'} &=& \dfrac{f^{}_\pi}{\st}\(\cos\theta\dfrac{f^{}_1}{f^{}_\pi}
- \s \sin\theta\dfrac{f^{}_8}{f^{}_\pi}\) \simeq 31.76~\MeV ~.~~
\eea
Using these we obtain
\bea
\dfrac{\cB(\tau\to\mu\eta)}{\cB(\tau\to\mu\phi)} \sim 0.60 \,, \quad
\dfrac{\cB(\tau\to\mu\eta')}{\cB(\tau\to\mu\phi)} \sim 1.9 \times 10^{-2} \,.
\eea

The current 90\% C.L. limits on these branching ratios are \cite{pdg}
\bea
\cB(\tau\to\mu\eta) &<& 6.5 \times 10^{-8} ~,~~ \nn\\
\cB(\tau\to\mu\eta') &<& 1.3 \times 10^{-7} ~,~~ \nn\\
\cB(\tau\to\mu\phi) &<& 8.4 \times 10^{-8} ~.~~
\eea
Of these decays, $\tau\to\mu\eta'$ is the least constraining. And
since $\tau\to\mu\phi$ and $\tau\to\mu\eta$ are of the same order, we
will use $\tau\to\mu\phi$ to constrain the coupling $\kappa$.  Using
$\cB(\tau\to\mu\phi) < 8.4 \times 10^{-8}$ \cite{taumuphiexp} and
$\La_\NP = 1$ TeV, we obtain the constraint
\bea
\l|\ka\r| < 0.019 ~.~~
\eea

\subsection{$\bs$-$\bsbar$ mixing}

As noted in Sec.~\ref{VBmodel}, the $VB$ model also generates four-quark
operators at tree level. In the mass basis, the operator of
Eq.~(\ref{4quark4lepton}) includes
\beq
\frac{(g_{qV}^{33})^2}{2 m^2_{V}} \sin^2 \theta_D \cos^2 \theta_D \, ({\bar s}_L \gamma^\mu b_L)\,({\bar s}_L \gamma_\mu b_L) ~.
\label{BsmixingVB}
\eeq
This generates
a contribution to $\bs$-$\bsbar$ mixing. In the SM, the same operator
is produced via a box diagram. Here we have
\beq
N C_{VLL}^{\rm SM} \, ({\bar s}_L \gamma^\mu b_L)\,({\bar s}_L \gamma_\mu b_L) ~,
\eeq
where
\bea
N & = & \frac{G_F^2 m_W^2}{16\pi^2} (V_{tb} V_{ts}^*)^2 \sim 10^{-11} \, \text{GeV}^{-2} ~, \nn\\
C_{VLL}^{\rm SM} & = & \eta_{B_s} x_t \left[ 1 + \frac{9}{1-x_t} - \frac{6}{(1-x_t)^2} -\frac{6 x_t^2 \ln x_t}{(1-x_t)^3} \right] ~.
\eea
In the above, $x_t \equiv m_t^2/m_W^2$ and $\eta_{B_s} = 0.551$ is the
QCD correction \cite{Buchalla:1995vs}.  The SM and NP contributions
can be combined. We define
\beq
N C_{VLL} \equiv N C_{VLL}^{\rm SM} + \frac{(g_{qV}^{33})^2}{2 m^2_{V}} \sin^2 \theta_D \cos^2 \theta_D ~.
\eeq
The mass difference in the $B_s$ system is then given by
\beq
 \Delta M_s = \frac{2}{3} m_{B_s} f_{B_s}^2 \hat B_{B_s} |N C_{VLL} | ~.
\eeq
Taking $f_{B_s}\sqrt{\hat B_{B_s}} = (266 \pm 18)$ MeV \cite{Aoki:2013ldr,Aoki:2016frl},
$V_{tb} V_{ts}^* = -0.0405 \pm 0.0012$ \cite{pdg,Charles:2015gya}, and $\overline{m}_t
= 160$ GeV \cite{pdg,Chetyrkin:2000yt}, we find the SM prediction
\beq
\Delta M_s^{\rm SM} = (17.4 \pm 2.6)~{\rm ps}^{-1} ~.
\eeq
This is to be compared with the experimental measurement \cite{HFAG}
\beq
\Delta M_s = (17.757 \pm 0.021)~{\rm ps}^{-1} ~.
\eeq
As we will see in the next section, the constraint on the $VB$ model from $\bs$-$\bsbar$
mixing is extremely stringent.

\subsection{$\tau \to 3\mu$}
\label{tau3mu}

Finally, the $VB$ model also produces four-lepton operators at tree
level.  In the mass basis, the Lagrangian of Eq.~(\ref{4quark4lepton})
includes the operator
\beq
-\frac{(g_{\ell V}^{33})^2}{2m^2_{V}} \sin^3 \theta_L \cos \theta_L \, ({\bar \mu}_L \gamma^\mu
\tau_L) \, ({\bar \mu}_L \gamma_\mu \mu_L) ~,
\label{tau3muamp}
\eeq
which generates the decay $\tau\to 3 \mu$.  As this is a
lepton-flavor-violating decay, it can arise only due to NP.  The
decay rate for $\tau \to 3 \mu$ is then given by
\beq
\cB(\tau^-\to\mu^-\mu^+\mu^-) = X \frac{(g_{\ell V}^{33})^4}{16m^4_V}\frac{m^5_\tau\tau_\tau}{192\pi
^3}\sin^6 \theta_L \cos^2 \theta_L ~,
\eeq
where $X$ is a suppression factor due to the non-zero muon mass. In
terms of $\eta_\mu = m_\mu/m_\tau$, it is given by
\bea
X &=& 12\int\limits^{(1 - \eta_\mu)^2}_{4\eta^2_\mu}\frac{dx}{\sqrt{x}}(x - 2\eta^2_
\mu)(1 + \eta^2_\mu - x)\sqrt{(x - 4\eta^2_\mu)(1 - 2(x + \eta^2_\mu) + (x + \eta^2_
\mu)^2)} \nl
&\approx& 0.94 ~.
\eea

At present, the branching ratio for $\tau^-\to\mu^-\mu^+\mu^-$ has
only an experimental upper bound \cite{tau23muexp}:
\beq
\cB(\tau^-\to\mu^-\mu^+\mu^-) < 2.1 \times 10^{-8} ~ {\rm at~90\%~C.L.}~
\label{tau3muexp}
\eeq
This then puts a constraint on $\theta_L$ in the $VB$ model,
  which, as we will see in the next section, is quite strong.

\section{Models: allowed parameter space}
\label{parameterspace}

\begin{figure}[t!]
\begin{center}
\includegraphics[viewport=0 0 360 400, width=16em]{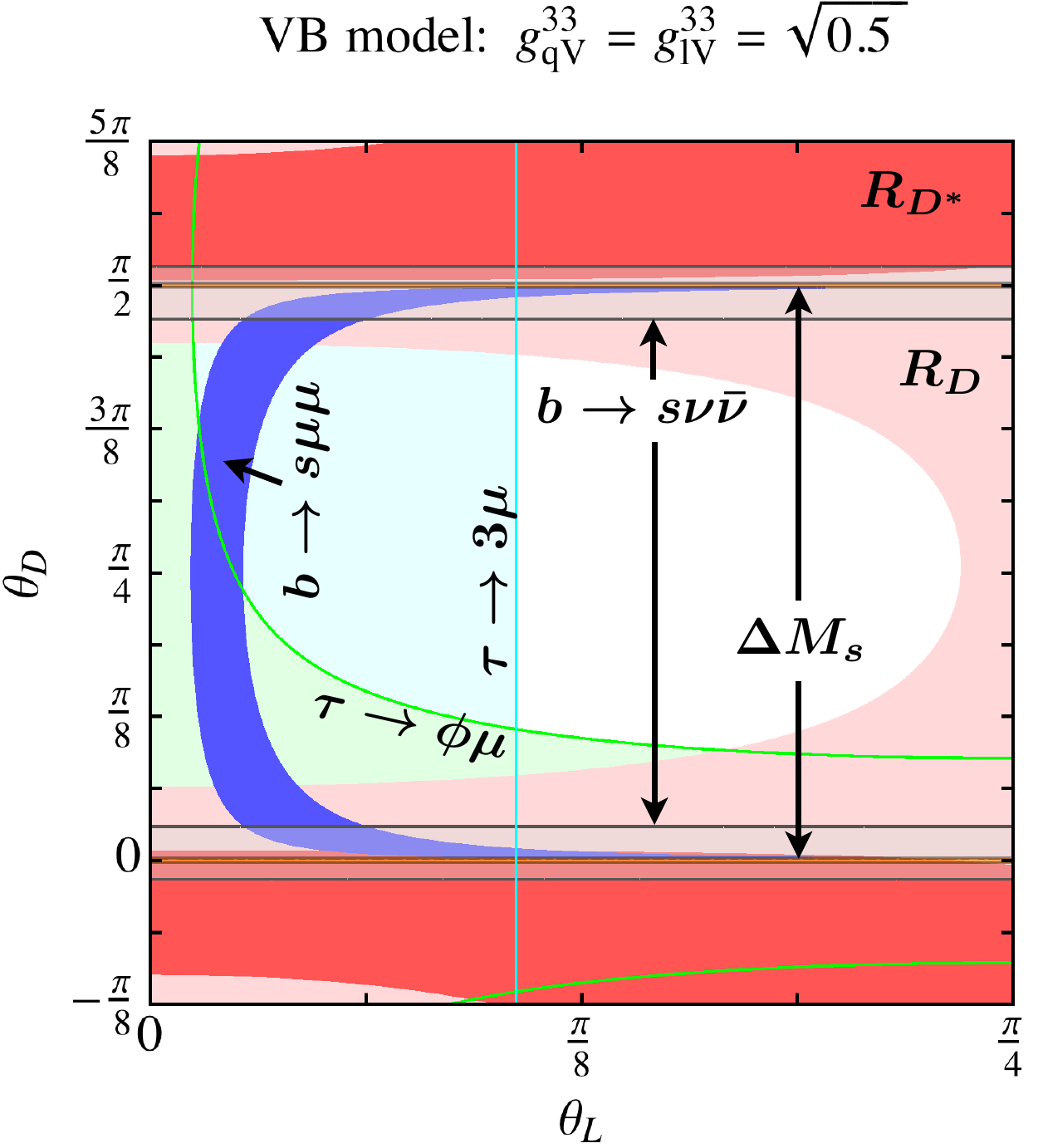}~~~~
\includegraphics[viewport=0 0 360 400, width=16em]{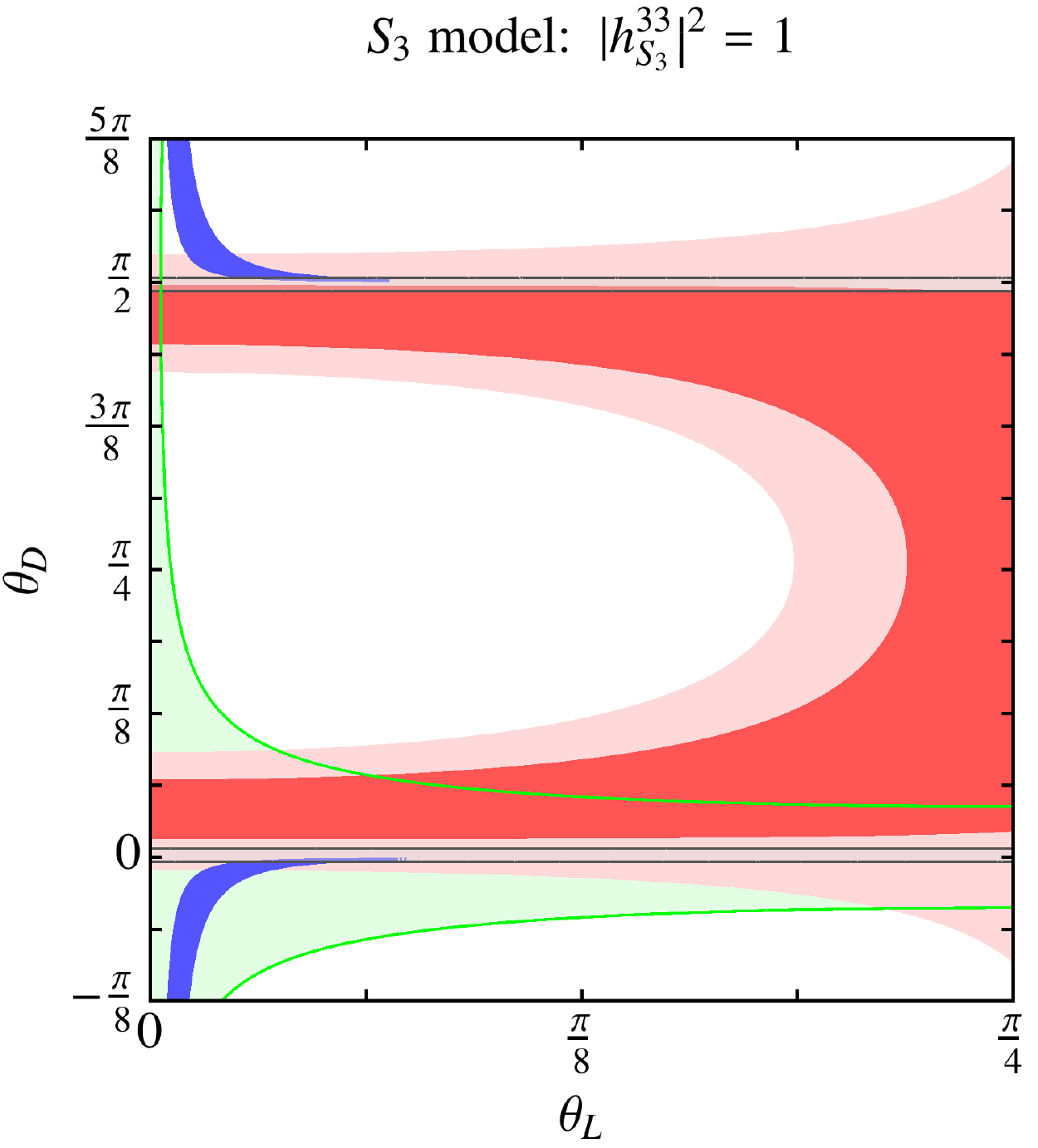} \\[1em]
\includegraphics[viewport=0 0 360 400, width=16em]{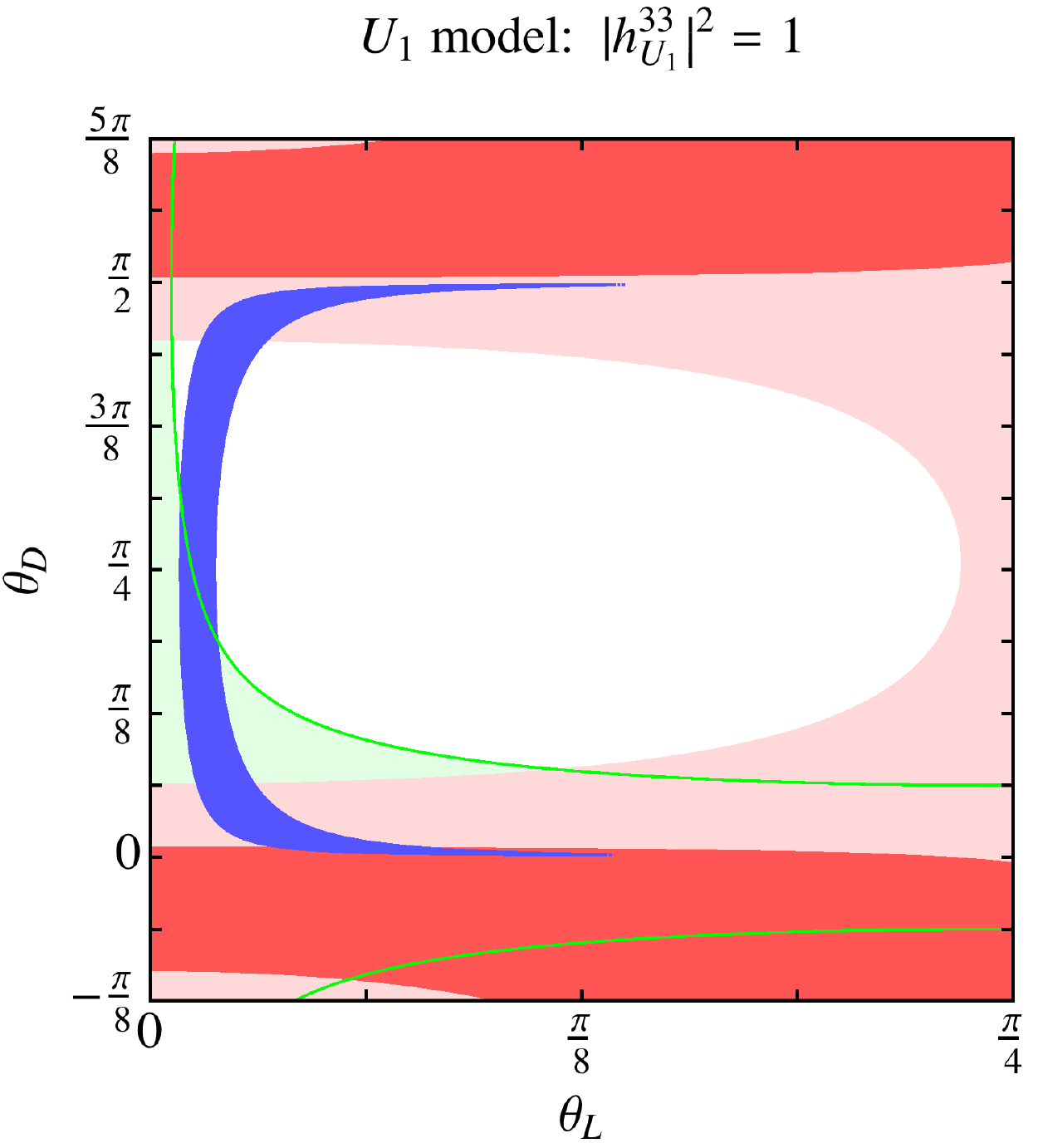}~~~~
\includegraphics[viewport=0 0 360 400, width=16em]{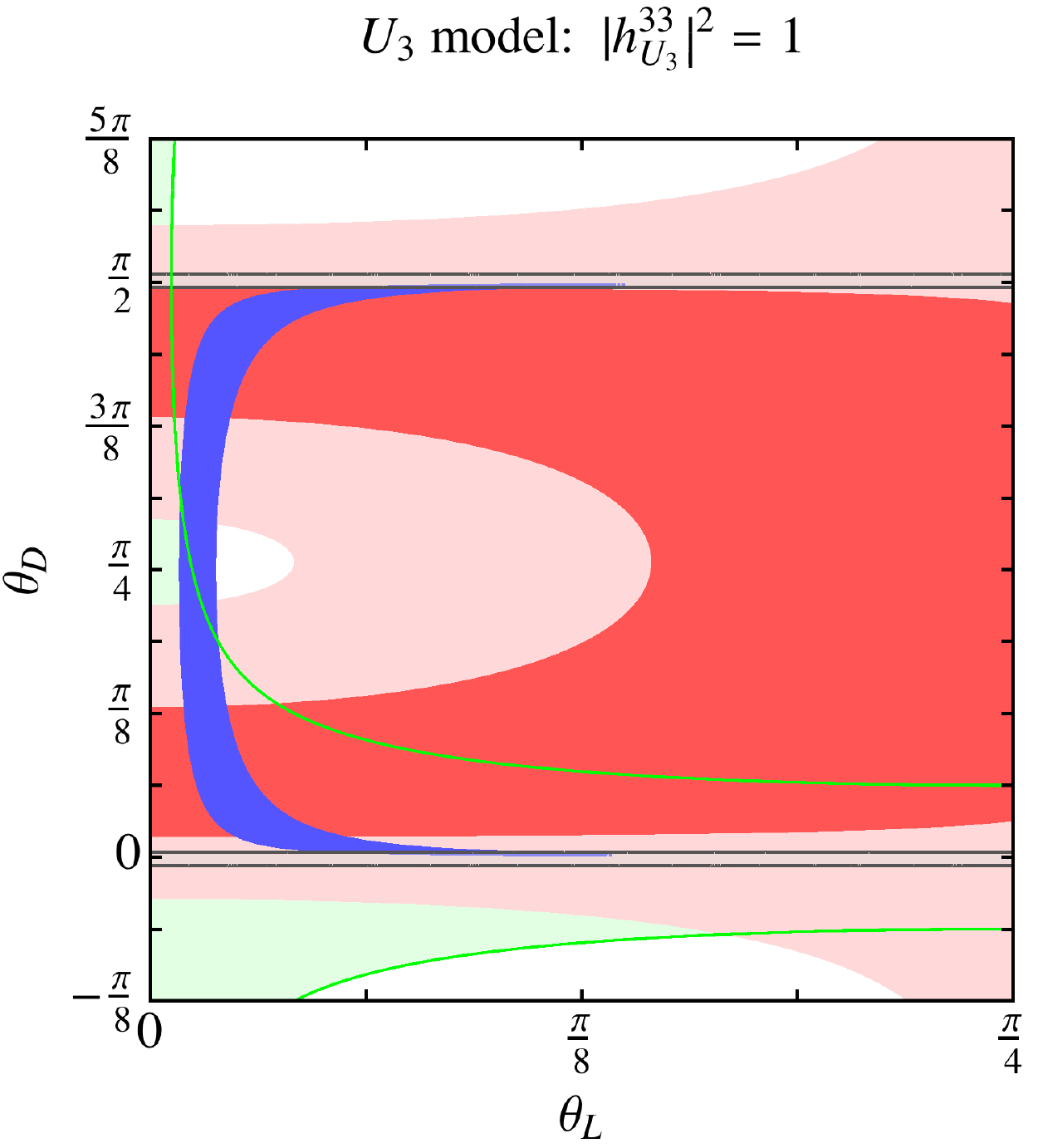}
\caption{Allowed regions in the $(\theta_L, \theta_D)$ plane for the
  $VB$, $S_3$, $U_1$, and $U_3$ models. We have fixed the NP scale as
  $\Lambda_\NP = 1\,\text{TeV}$. In each model, the third-generation
  coupling is taken as $2g^{33}_{qV} g^{33}_{\ell V} =\l|
  h_{U_1}^{33}\r|^2 =\l|h^{33}_{U_3}\r|^2 =\l| h_{S_3}^{33}\r|^2 =1$.
  The $R_D$, $R_{D^*}$ and $R_K$ (along with the $\bsll$ data)
  anomalies can be explained in the shaded regions colored in pink,
  red, and blue, respectively. The regions bounded by the gray,
    green, cyan, and orange lines are allowed from the measurements of
    $b\to s\nu\bar\nu$, $\tau\to\mu\phi$, $\tau\to 3\mu$, and $\Delta
    M_s$, respectively.  The last two observables are applicable only
    in the $VB$ model.}
\label{Fig:allowedregions}
\end{center}
\end{figure}

Taking into account all the experimental constraints described in
Sec.~\ref{constraints}, we find the allowed parameter space in the
four NP models.  We assume $\Lambda_\NP = 1$ TeV, and take the
third-generation coupling to be $2g^{33}_{qV} g^{33}_{\ell V} =\l|
h_{U_1}^{33}\r|^2 =\l|h^{33}_{U_3}\r|^2 =\l| h_{S_3}^{33}\r|^2 =1$.
For the $VB$ model, we take $g^{33}_{qV} = g^{33}_{\ell V}$.
(In the next section we vary $g^{33}_{qV}$ and $g^{33}_{\ell V}$.)
In Fig.~\ref{Fig:allowedregions}, the constraints in the $(\theta_L,
\theta_D)$ plane are shown for the $VB$, $U_1$, $U_3$ and $S_3$
models. These are presented only for $\theta_L \ge 0$; the space is
symmetric under $\theta_L \to -\theta_L$.

For all four models, the flavor anomalies $R_D$, $R_{D^*}$ and $R_K$
can be explained in the shaded regions colored in pink, red and
blue, respectively.  The gray shaded region is allowed from $\bar B
\to K^{(*)} \nu\bar\nu$ at 90\% C.L.  The region bounded by the green
lines is consistent with the 90\% C.L. upper limit on the branching
ratio of $\tau\to\mu\phi$. For the $VB$ model, there are additional
constraints coming from $\bs$-$\bsbar$ mixing and $\tau\to 3\mu$. For
the $\tau\to 3\mu$ constraint, the region to the left of the cyan line is allowed.
The $\bs$-$\bsbar$ mixing constraint is shown in the orange region,
which is extremely narrow near $\theta_D = 0, \pi/2$.

Based on this figure, one can make two observations:
\begin{itemize}

\item There are only two regions in parameter space where the
  constraints from $R_D$, $R_{D^*}$, $R_K$ and $\bar B \to K^{(*)}
  \nu\bar\nu$ (if applicable) might overlap. These are roughly around
  $\pi/16 \lesssim \theta_L \lesssim \pi/8$, with $\theta_D$ near 0
  (region 1) or $\pi/2$ (region 2). However, the additional constraint
  from $\tau \to\phi\mu$ distinguishes the two regions. That is, while
  region 1 satisfies the $\tau \to\phi\mu$ constraint, region 2 does
  not, and is therefore excluded. Henceforth, we focus only on region
  1.

\item For the $VB$ model, the constraint from $\bs$-$\bsbar$ mixing has
  the same shape as that from $\bar B \to K^{(*)} \nu\bar\nu$. They
  are both independent of $\theta_L$, and so bound only $\theta_D$.
  However, we see that the $\bs$-$\bsbar$ mixing constraint is
    much more stringent than that from $\bar B \to K^{(*)}\nu\bar\nu$.
    For $g^{33}_{qV}=g^{33}_{\ell V}= 1/\s$, one has $|\theta_D|
    \ll 1$, so that it is somewhat difficult from this figure to
    determine if this region is consistent with the others.

\end{itemize}

\begin{figure}[t]
\begin{center}
\includegraphics[viewport=0 0 360 395, width=16em]{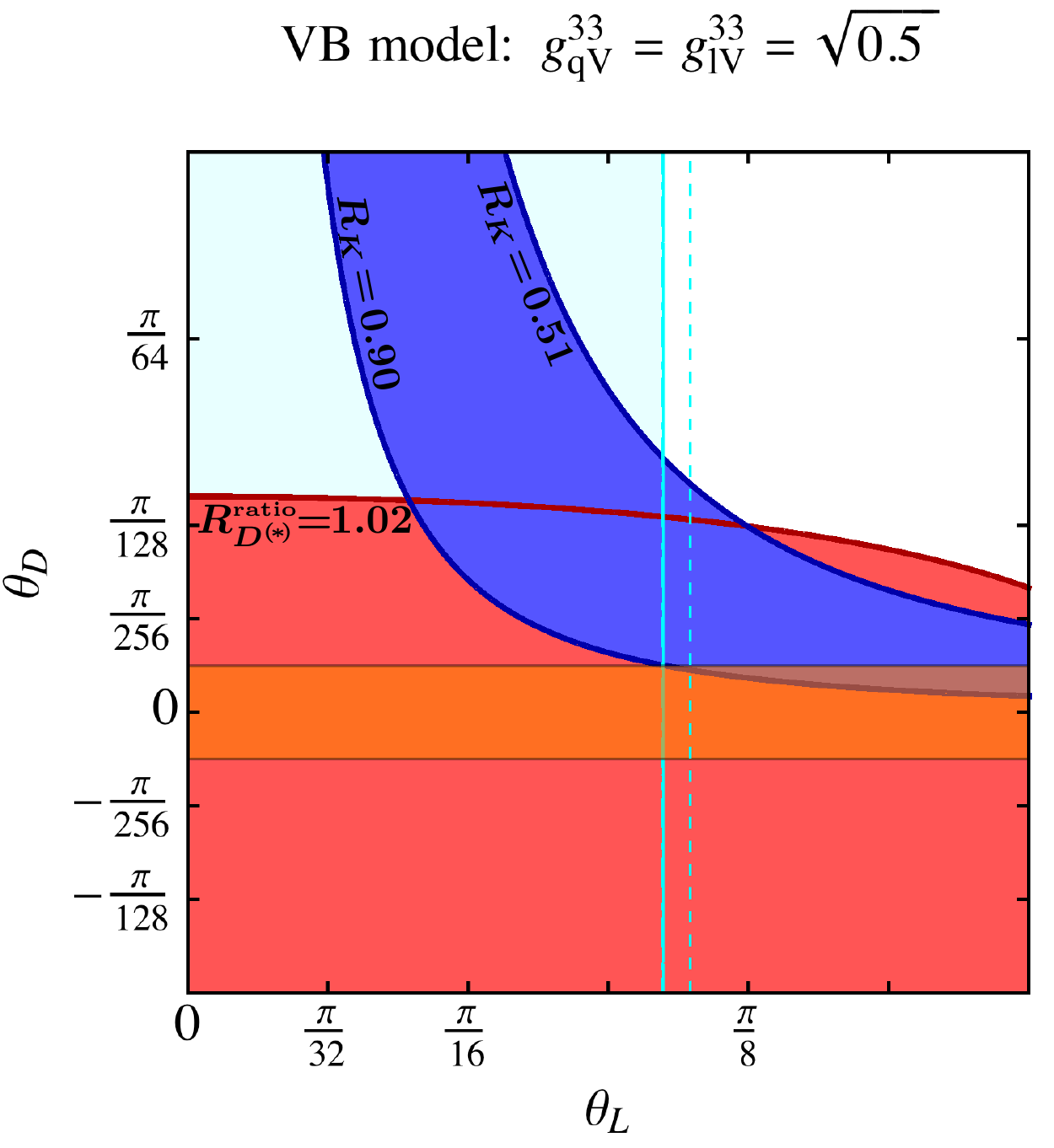}~~~~
\includegraphics[viewport=0 0 360 395, width=16em]{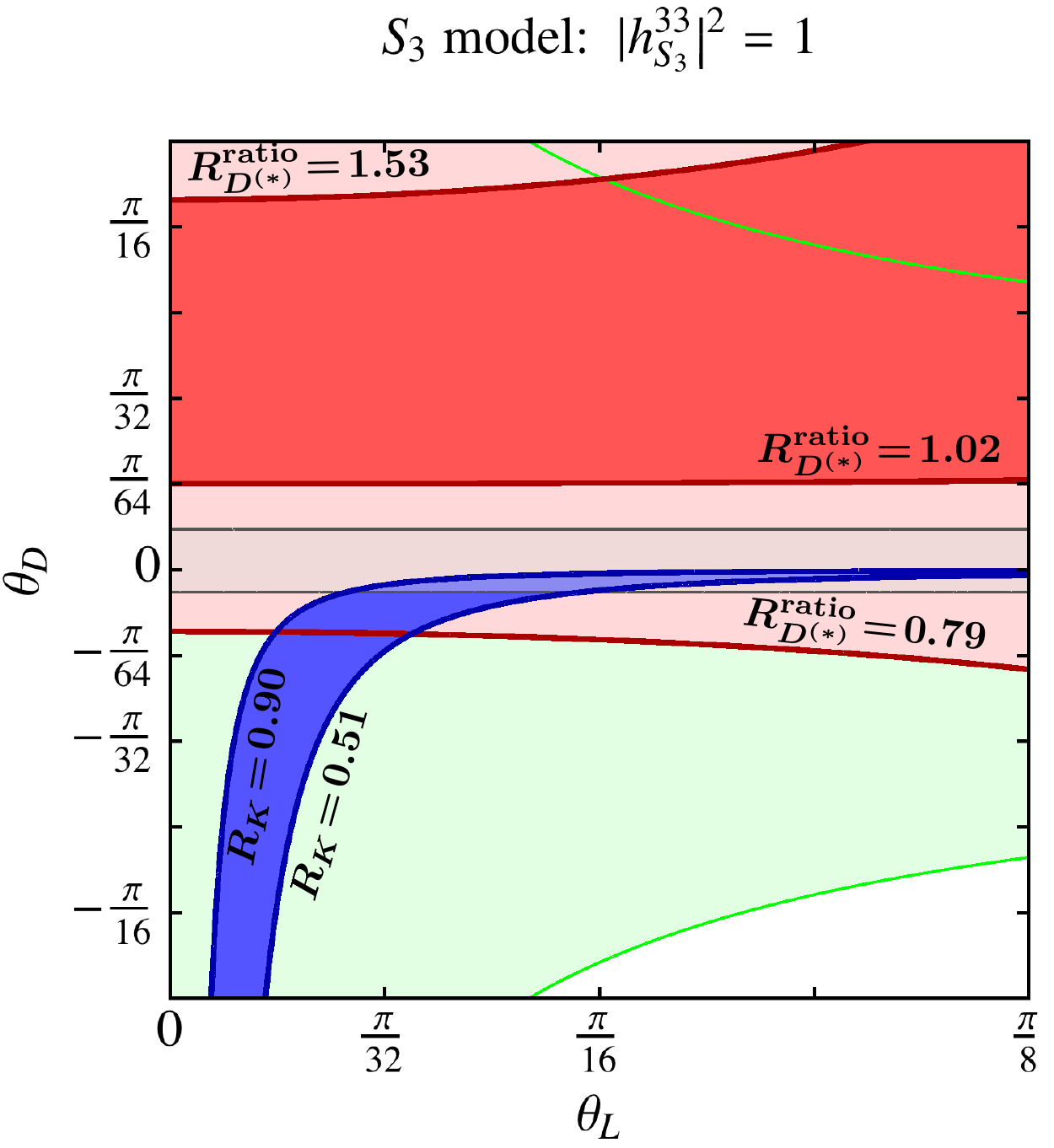} \\[1em]
\includegraphics[viewport=0 0 360 395, width=16em]{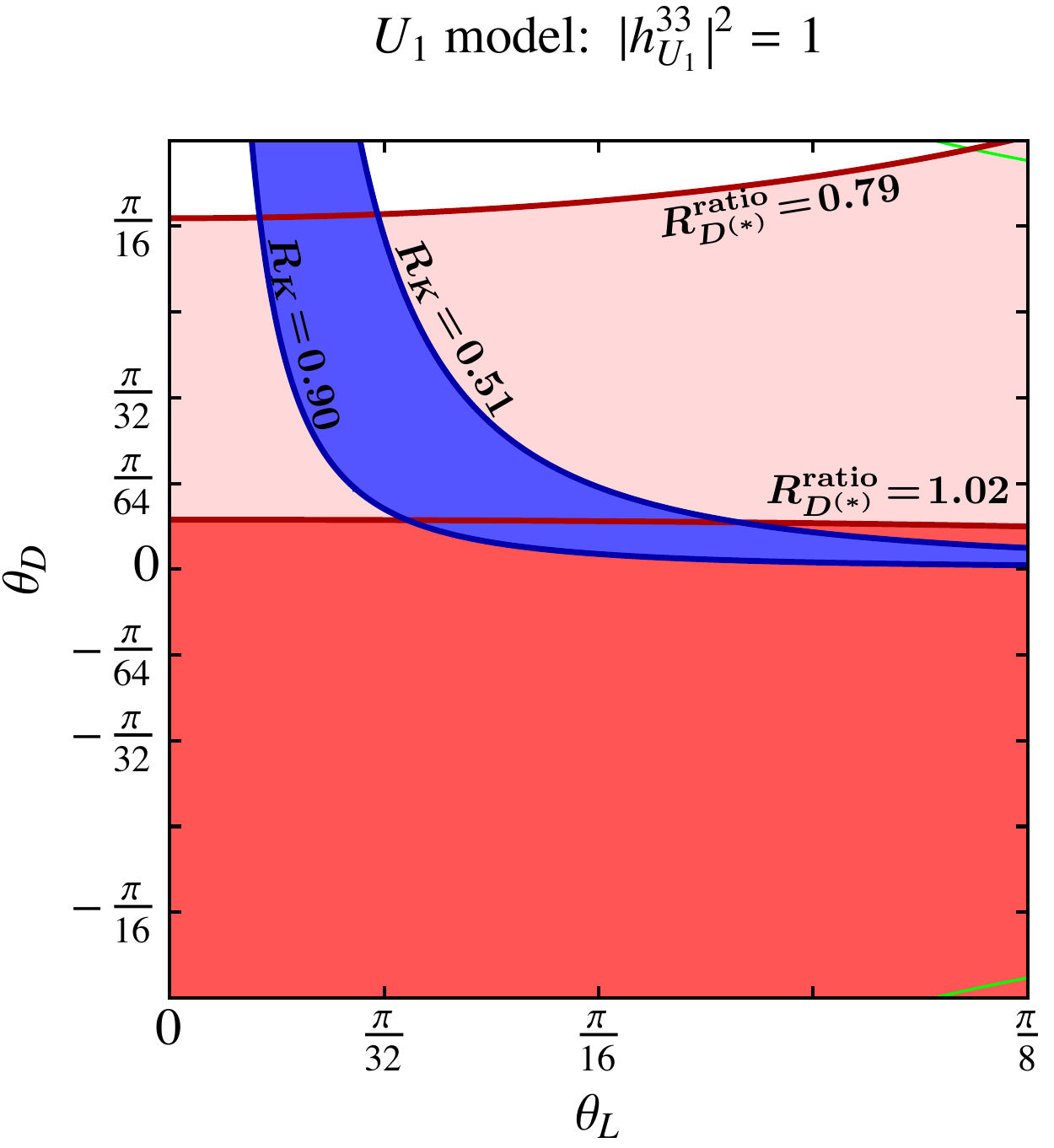}~~~~
\includegraphics[viewport=0 0 360 395, width=16em]{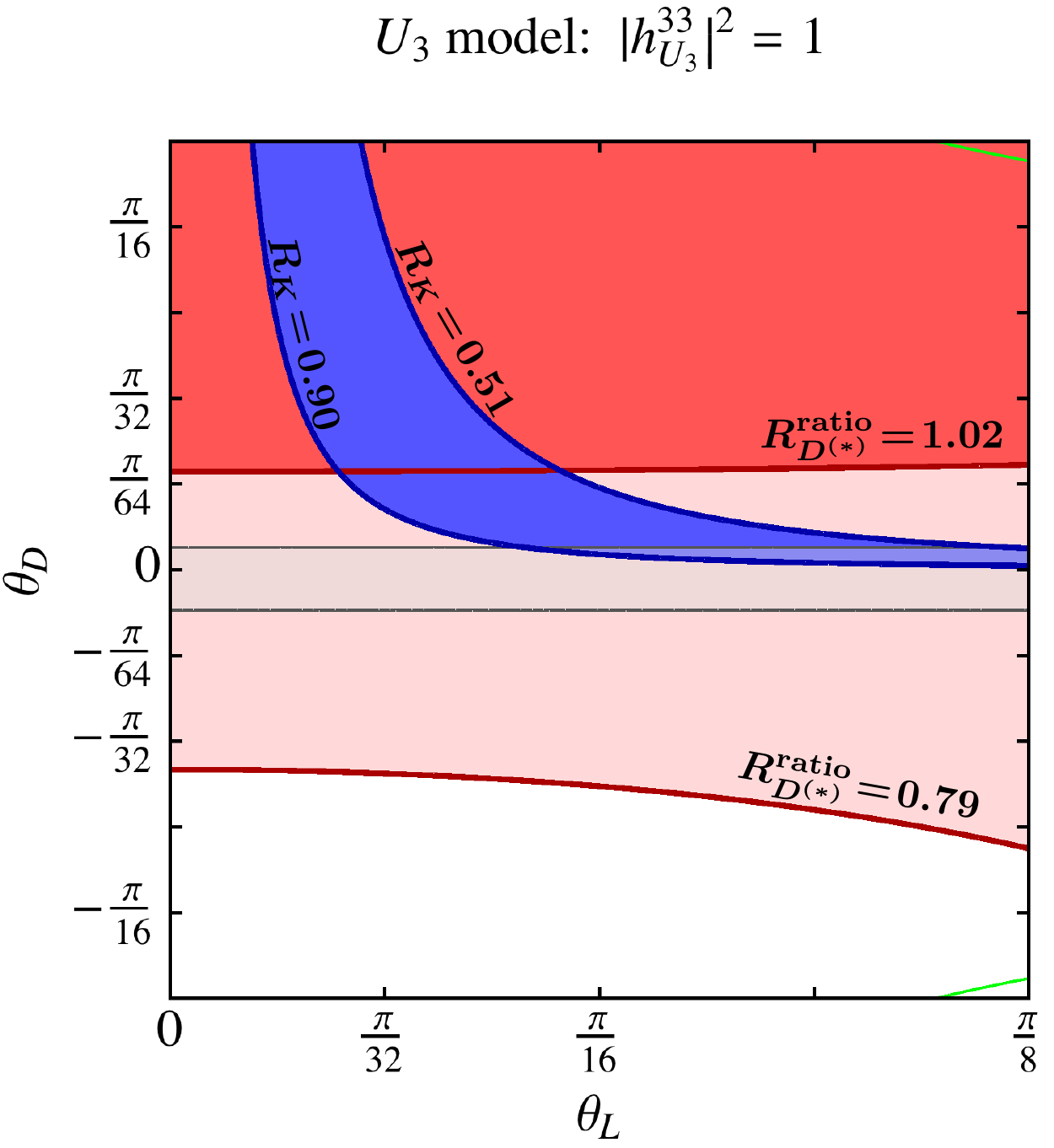}
\caption{Magnified figures of Fig.~1. The color legends are the same
  as the previous figures.  The values of the contours for $R_X^{\rm
    NP+SM}/R_X^{\rm SM}$ ($X=K, D, D^*$) are indicated. The $R_D$
    constraint (in pink) is satisfied for the entire region of the plot in
    $VB$ and hence omitted. }
\label{Fig:magnifiedregions}
\end{center}
\end{figure}

In order to obtain more information, in
Fig.~\ref{Fig:magnifiedregions}, we show the constraints in region~1
of the $(\theta_L, \theta_D)$ plane for the $VB$, $S_3$, $U_1$, and
$U_3$ models. In the figures, we indicate the values of the contours
for the flavor anomalies, that is, $R_X^{\rm NP+SM}/R_X^{\rm SM}$ for
$X=K$ and $D^{(*)}$.  From this figure, we can see that
\begin{itemize}

\item For the $S_3$ model, the $R_{D^*}$ region does not overlap with
  the $R_K$ or $\bar B \to K^{(*)} \nu\bar\nu$ regions.  And for the
  $U_3$ model, the $R_{D^*}$ and $\bar B \to K^{(*)} \nu\bar\nu$
  regions do not overlap.  Therefore, $S_3$ and $U_3$ with the flavor
  mixing structure of Eq.~(\ref{XYdefs}) are excluded\footnote{ A
    different type of the mixing for the $U_3$ LQ is discussed in
    Ref.~\cite{U3LQ}}.

\item
  For the $U_1$ model, if $\theta_D \le 0.028$, there is a
    region where all the constraints overlap, and so this model is
    allowed.  On the other hand, the $VB$ model is on the edge of
    exclusion -- the boundaries of the $\Delta M_s$, $b \to s\mu\mu$,
    and $\tau\to 3\mu$ constraints are touching, but just barely.
    This also implies that the $VB$ model only allows limited values
    of these observables. From this figure we see that $\tau\to 3\mu$
    is a critical process for the $VB$ model. Therefore, in addition
    to the 90\% C.L. upper bound on $\cB(\tau^-\to\mu^-\mu^+\mu^-)$
    from Eq.~(\ref{tau3muexp}) shown in the figure (solid cyan line),
    we superpose an estimated $3\sigma$ upper limit on this branching
    ratio (dashed cyan line).

\item
  Comparing the $VB$ and $U_1$ models, the constraints from the
  flavor anomalies ($R_K$, $R_{D^{(*)}}$) are similar (for $\theta_D$
  near 0). But the additional constraints from $\Delta M_s$ and $\tau
    \to 3\mu$ put $VB$ on the verge of exclusion. The limited amount
    of parameter space available to $VB$ increases its predictive power.
    Specifically, while the allowed region for the $U_1$ model includes
    $0.11 \le \theta_L\le 0.73$ and $0.001 \le \theta_D \le 0.028$, for the
    $VB$ model $(\theta_L, \theta_D)$ is limited to be $\simeq (0.333,0.006)$.

\end{itemize}

The $VB$ and $U_1$ models are therefore the candidates to
simultaneously explain the $R_K$ and $R_{D^{(*)}}$ puzzles in the case
where the NP couples predominantly to the third-generation fermions.

\section{Predictions}

We have now established that the $VB$ and $U_1$ models are candidates
to explain the present discrepancies with the SM in $\bsmumu$,
$R_{D^{(*)}}$ and $R_K$. The main question we wish to address in this
paper is: is there any way of distinguishing the two models? There are
two handles that can potentially accomplish this. First, the $VB$
model contributes to four-lepton and four-quark operators, and hence
to processes such as $\tau \to 3\mu$ and $\bs$-$\bsbar$ mixing, while
the $U_1$ model does not\footnote{In this paper we perform the
  analysis at tree level. Radiative corrections to four-lepton
  operators have been considered in Ref.\ \cite{ParadisiEFT} within an
  EFT framework.  However, as with all EFT analyses, the results do
  not necessarily apply to all models. To obtain the proper result, a
  more complete analysis must be done within each individual
  model.}. Second, due to additional constraints, the allowed region
in $(\theta_L,\theta_D)$ space is essentially a single point for $VB$, while it
  is much larger for $U_1$. Below we explore the predictions of the
two models for various processes. As we will see, it is potentially
possible to distinguish the $VB$ and $U_1$ models.

\subsection{Processes}
\label{Pred_processes}

\subsubsection{$R_{D^{(*)}}$}

The $3\sigma$ allowed ranges of $R_{D^{(*)}}^{\rm ratio}$ are given in
Eq.~(\ref{RDconstraint}). At present, large deviations from the SM are
allowed (up to 79\% and 53\% for $R_{D}$ and $R_{D^{*}}$,
respectively). On the other hand, from Fig.~2, we see that the $VB$
and $U_1$ models are allowed only if $\theta_D$ is very small.  This
means that such large deviations in $R_{D^{(*)}}$ from the SM are not
favored, as these are inconsistent with the $R_K$ anomaly. The models
predict
\bea
VB &~~:~~& R_{D^{(*)}}^{\rm ratio}\, \simeq 1.04 ~, \nn\\
U_1 &~~:~~& 1.02 \le R_{D^{(*)}}^{\rm ratio} \le 1.05 ~.
\eea
Thus, even if $R_{D^{(*)}}$ is measured with greater precision, it
will probably not be possible to distinguish the $VB$ and $U_1$
models. However, if the measurements confirm large deviations from the
SM, both models will be ruled out.

\subsubsection{$R_K$}

The situation is different for $R_K$. Using Eq.~(\ref{RKexp}), its
allowed $3\sigma$ range is $0.498 \le R_K \le 1.036$.  The models
predict~\cite{flavio}
\bea
VB &~~:~~& R_K\, \simeq 0.90 ~, \nn\\
U_1 &~~:~~& 0.51 \le R_K \le 0.90 ~.
\label{RKcoup1}
\eea
We therefore see that the $U_1$ model can accomodate smaller values of
$R_K$ than can the $VB$ model. This is due to the fact that its
allowed $(\theta_L,\theta_D)$ region includes larger values of
$\theta_L$. Thus, if future measurements of $R_K$ find it to be
less than 0.90 at higher than 90\% C.L., this would point clearly
to $U_1$ (and exclude $VB$).

\subsubsection{$\tau \to 3\mu$}

This decay is particularly interesting because only the $VB$ model
contributes to it.  The present experimental bound is
$\cB(\tau^-\to\mu^-\mu^+\mu^-) < 2.1 \times 10^{-8}$ at 90\%
C.L.\ [Eq.~(\ref{tau3muexp})]. Belle II expects to reduce this limit
to $< 10^{-10}$ \cite{Shintalk}. The reach of LHCb is somewhat weaker,
$< 10^{-9}$ \cite{LHCbLetterofIntent}.

Now, the amplitude for $\tau \to 3\mu$ depends only on $\theta_L$
[Eq.~(\ref{tau3muamp})]. In Fig.~\ref{Fig:magnifiedregions}, we see that
  the allowed value of $\theta_L$ corresponds to the present
  experimental bound. That is, $VB$ predicts
\beq
\cB(\tau^-\to\mu^-\mu^+\mu^-)\, \simeq 2.1 \times 10^{-8} ~.
\label{tau3mucoup1}
\eeq
Thus, the $VB$ model predicts that $\tau \to 3\mu$ should be observed
at both LHCb and Belle II. This is a smoking-gun signal for the model.

\subsubsection{$B \to K^{(*)} \mu \tau$}

The BaBar Collaboration obtained an experimental bound of $\cB(B^+ \to
K^+ \mu^\pm \tau^\mp) < 4.8 \times 10^{-5}$ at 90\%
C.L.\ \cite{BKmutauexp}. Belle II will collect 100 times more data
than BaBar, and this will allow it to measure $\cB(B^+ \to K^+ \mu^\pm
\tau^\mp)$ to a level of $5 \times 10^{-7}$ \cite{Belleprivcomm}.

The models predict~\cite{flavio}
\bea
VB &~~:~~& \cB(B \to K^{(*)} \mu \tau)\, \simeq 4.0 \times 10^{-10}~, \nn\\
U_1 &~~:~~& 6.8 \times 10^{-11} \le \cB(B \to K^{(*)} \mu \tau) \le 2.1 \times 10^{-8} ~.
\label{pred_BKmutau}
\eea
Neither model can produce $\cB(B \to K^{(*)} \mu \tau)$ sufficiently
large that it can be observed at Belle II.

\subsubsection{$B \to K^{(*)} \tau^+ \tau^-$}

The BaBar Collaboration recently put a limit of $\cB(B^+ \to K^+
\tau^+ \tau^-) < 2.25 \times 10^{-3}$ at 90\%
C.L.\ \cite{BaBarBKtautau}. Belle II will be able to improve on this,
but because there are two $\tau$'s in the final state, the expected
reach is only $\sim 2 \times 10^{-4}$ \cite{Belleprivcomm}.

To measure and calculate the branching ratio of $B \to K^{(*)} \tau^+
\tau^-$, we need to deal with charmonium resonances. In analogy with
$B \to K^{(*)} \mu^+ \mu^-$, we take $q^2 > 15\,\text{GeV}^2$ for
integration and obtain the partial branching ratio by using {\tt
  flavio}~\cite{flavio}:
\bea
VB &~~:~~& \cB(B \to K^{(*)} \tau^+ \tau^-)\, \simeq 4.4 \times 10^{-8} ~, \nn\\
U_1 &~~:~~& 7.6 \times 10^{-10} \le \cB(B \to K^{(*)} \tau^+ \tau^-) \le 1.5 \times 10^{-6} ~.
\eea
The values of $\cB(B \to K^{(*)} \tau^+ \tau^-)$ possible in both models
are at least two orders of magnitude smaller than the estimated reach of
Belle II. This decay can therefore not be used as a signal of the $VB$
and/or $U_1$ models.

\subsubsection{$\bs \to \mu\tau$, $\bs \to \tau^+ \tau^-$}

At present, LHCb is working on measuring these two decays, which are
difficult due to the presence of $\tau$'s in the final state. However,
no estimates of the reach are available \cite{LHCbprivcomm}.  (At
Belle~II, a rough estimate for $\bs \to \tau^+ \tau^-$ could be $\sim
2 \times 10^{-3}$ with $50\,\text{ab}^{-1}$ of data, obtained by
rescaling the present data at Belle.)

For $\bs \to \mu\tau$, the models predict
\bea
VB &~~:~~& \cB(\bs \to \mu\tau)\, \simeq 6.7 \times 10^{-9} ~, \nn\\
U_1 &~~:~~& 1.1 \times 10^{-9} \le \cB(\bs \to \mu\tau) \le 3.6 \times 10^{-7} ~,
\label{pred_Bsmutau}
\eea
while for $\bs \to \tau^+ \tau^-$ we have
\bea
VB &~~:~~& \cB(\bs \to \tau^+ \tau^-)\, \simeq 2.4 \times 10^{-7}~, \nn\\
U_1 &~~:~~& 5.8 \times 10^{-18} \le \cB(\bs \to \tau^+ \tau^-) \le 6.7 \times 10^{-6} ~.
\eea
For $\bs \to \mu\tau$, if the branching ratio were measured to be between
$6.7 \times 10^{-9}$ and $3.6 \times 10^{-7}$,
  this would point to the $U_1$ model.  However, it is unlikely that
such a small branching ratio is measurable. Similarly, if $\cB(\bs \to
  \tau^+ \tau^-)$ were found to be in the range $2.4 \times
  10^{-7}$--$6.7 \times 10^{-6}$, this would indicate $U_1$.
However, here too it is not clear that such a small branching ratio is
measurable.

\subsubsection{$\Upsilon \to \mu \tau$}

Finally, we turn to $\Upsilon \to \mu \tau$.  This
lepton-flavor-violating decay has been overlooked in previous
analyses, but it is potentially an important process to consider
{\footnote{ Quark flavor violating quarkonium decays were considered
    in Ref.~\cite{Datta:1998yq}.}. At the fermion level, this decay is
  $b{\bar b} \to \mu \tau$, and it can receive contributions from both
  the $VB$ and $U_1$ models.  Note that this process has a pattern of
  mixing different from the above processes, and thus the models
  provide unique predictions.

In the past, the BaBar \cite{BaBarUps} and CLEO \cite{CLEOUps}
Collaborations have studied lepton flavor violation in narrow
$\Upsilon(nS) (n = 1,2,3)$ decays. The strongest limits come from
BaBar \cite{BaBarUps}, which put an upper limit on
$\cB(\Upsilon(2S,3S) \to \mu \tau)$ of a few times 10$^{-6}$. This was
obtained using 13.6 fb$^{-1}$ and 26.8 fb$^{-1}$ of the BaBar dataset
on the $\Upsilon(2S)$ and $\Upsilon(3S)$, respectively. Belle II is
expected to collect a few hundred fb$^{-1}$ of data on the
$\Upsilon(3S)$ \cite{Belleprivcomm}. A precise estimate of the
sensitivity to $\Upsilon(3S) \to \mu \tau$ will require a dedicated
study. However, given the order-of-magnitude increase in luminosity at
Belle II compared to BaBar, we expect roughly an order-of-magnitude
improvement in the sensitivity. That is, a reach of about 10$^{-7}$
for $\cB(\Upsilon(3S) \to \mu \tau)$ at Belle II is not
unreasonable. These decays may also be studied at LHCb, but we are not
aware of the LHCb reach for these processes.

In the SM, the LFV decay $\Upsilon(nS)\to\ell^-\ell'^+$, where $\ell$
and $\ell'$ represent leptons of different flavor, is highly
suppressed. On the other hand, in the $VB$ and $U_1$ models,
$\Upsilon(nS)\to\mu^-\tau^+$ receives significant contributions.
Assuming the NP is purely left-handed, the decay rate for this process
is given by
\bea
\label{Eq:LFVUps}
\Ga(\Upsilon(nS)\to\mu^-\tau^+) &=& \frac{m^3_{\Upsilon(nS)} f^2_{\Upsilon(nS)}}{48\pi}
(1 - \eta^2_\tau)(2 - \eta^2_\tau - \eta^4_\tau)|\kappa|^2 ~,~~
\eea
where $\eta_\tau = m_\tau/m_{\Upsilon(nS)}$ and $\kappa$ contains the
coupling corresponding to the transition $b{\bar b} \to\tau\mu$. In
the $VB$ and $U_1$ models we have
\beq
\kappa = -\frac{g_1 + g_2}{2\Lambda^2_\NP}X^{33}Y^{32}
= \frac{g_1 + g_2}{2\Lambda^2_\NP}\cos^2\theta_D\cos\theta_L\sin\theta_L ~.~~
\label{Eq:UpsWC}
\eeq
The decay constant $f_{\Upsilon(nS)}$ can be found using the
electromagnetic decay $\Upsilon(nS) \to \ell^-\ell^+$,
which is unaffected by NP.
Its decay rate can be expressed as
\bea
\label{Eq:FCUps}
\Ga(\Upsilon(nS)\to\ell^-\ell^+) &=& \frac{4\pi\al^2}{27}\frac{f^2_
{\Upsilon(nS)}}{m_{\Upsilon(nS)}}(1 + 2\eta^2_{\ell(nS)})\sqrt{1 - 4
\eta^2_{\ell(nS)}} ~,~~
\eea
where $\eta_{\ell(nS)} = m_\ell/m_{\Upsilon(nS)}$.

We may now combine Eqs.~(\ref{Eq:LFVUps}), (\ref{Eq:UpsWC}) and
(\ref{Eq:FCUps}) to get predictions for the branching ratio of
$\Upsilon(nS)\to\mu\tau$ in the $VB$ and $U_1$ models. These are
\bea
VB &~~:~~& \cB(\Upsilon(1S) \to \mu \tau)\, \simeq 2.3 \times 10^{-9} ~, \nn\\
&&  \cB(\Upsilon(2S) \to \mu \tau)\, \simeq 2.3 \times 10^{-9} ~, \nn\\
&&  \cB(\Upsilon(3S) \to \mu \tau)\, \simeq 3.0 \times 10^{-9} ~, \nn\\[0.5em]
U_1 &~~:~~& 1.1 \times 10^{-9} \le \cB(\Upsilon(1S) \to \mu \tau) \le 2.4 \times 10^{-8} ~, \nn\\
&& 1.2 \times 10^{-9} \le \cB(\Upsilon(2S) \to \mu \tau) \le 2.4 \times 10^{-8} ~, \nn\\
&& 1.5 \times 10^{-9} \le \cB(\Upsilon(3S) \to \mu \tau) \le 3.2 \times 10^{-8} ~.
\eea
The $VB$ model predicts a branching ratio of $O(10^{-9})$, while
it can be $O(10^{-8})$ in the $U_1$ model. Therefore this mode can
potentially allow us to distinguish between the two models. However,
even the upper limit predicted by the $U_1$ model seems to be out of
reach of Belle II, according to our estimate of its reach. On the
other hand, perhaps Belle II or LHCb will in fact be sensitive to
branching ratios of $O(10^{-8})$. Or perhaps the NP coupling is bigger
than we have assumed (see Sec.~\ref{Couplingsvaried} below), resulting
in larger branching ratios. The point is that $\Upsilon \to \mu \tau$
decays may provide us with valuable information in identifying the
lepton-flavor-violating NP.

\subsubsection{Summary}

There are therefore three observables that can distinguish the $VB$
and $U_1$ models:
\begin{enumerate}

\item $\tau \to 3\mu$: $VB$ predicts
  $\cB(\tau^-\to\mu^-\mu^+\mu^-) \simeq 2.1 \times 10^{-8}$, its
  present upper limit ($U_1$ does not contribute to the decay). This
  implies that the LFV decay $\tau \to 3\mu$, which is absent in the
  SM, should be observed at both LHCb and Belle II. This is therefore
  a smoking-gun signal: it can occur only in the $VB$ model, and if
  the decay is not seen, the model would be ruled out.

\item $R_K$: The current $3\sigma$ range for $R_K$ is $0.498 \le R_K
  \le 1.036$. The $U_1$ model can accomodate smaller values of $R_K$,
  while the $VB$ model cannot. Specifically, if future measurements of
  $R_K$ find it to be less
    than 0.90 at higher than 90\% C.L., this would point to
    $U_1$ (and exclude $VB$).

\item $\Upsilon \to \mu \tau$: To date, the LFV decay $\Upsilon \to
  \mu \tau$ has been overlooked as a test of NP models in $B$ decays.
  Within the $VB$ model, $\cB(\Upsilon(nS) \to \mu \tau)$ is
  a few times $10^{-9}$, but in the $U_1$ model, it can reach a
  few times $10^{-8}$. Belle II should be able to measure
  $\cB(\Upsilon(3S) \to \mu \tau)$ down to $\sim 10^{-7}$. However,
  this is only a very rough estimate -- a detailed study is needed for
  a precise determination of the reach. It may be that, in fact, Belle
  II (or LHCb) will be able to observe branching ratios of
  $O(10^{-8})$. And if the decay $\Upsilon \to \mu \tau$ is observed,
  this will suggest the $U_1$ model.

\end{enumerate}
There are five other observables that receive contributions in the
$VB$ and $U_1$ models: $R_{D^{(*)}}$, $B \to K^{(*)} \mu \tau$, $B \to
K^{(*)} \tau^+ \tau^-$, $\bs \to \mu\tau$, $\bs \to \tau^+ \tau^-$.
However, either these observables cannot distinguish the two models,
or, if they can, the predicted branching ratios fall below the
expected reach of Belle II and LHCb.

\subsection{Varying the Couplings}
\label{Couplingsvaried}

Now, the results of the previous subsection have been found assuming
that $2 \, g^{33}_{qV} g^{33}_{\ell V} =\l| h_{U_1}^{33}\r|^2
=1$. However, there is nothing special about this value of the square
of the coupling (henceforth denoted coupling$^2$). This then raises
the question: if the coupling$^2$ is allowed to take different values,
how do the results of Sec.~\ref{Pred_processes} change? This is
examined in this subsection.

For each new value of the coupling$^2$, one must redo the analysis of
Sec.~\ref{parameterspace}, to determine the region in $(\theta_L,
\theta_D)$ parameter space allowed by the various experimental
constraints. That is, figures of the type in
Fig.~\ref{Fig:magnifiedregions} are produced. The following results
are found:
\begin{itemize}

\item For the $S_3$ and $U_3$ models, it is found that the $R_{D^*}$
  and $\bar B \to K^{(*)} \nu\bar\nu$ regions do not overlap, and this
  is independent of the value of coupling$^2$. $S_3$ and $U_3$ are
  therefore excluded.

\item For
  the $VB$ model, the constraints essentially come from three
  observables:
\begin{enumerate}

\item $\bs$-$\bsbar$ mixing ($\Delta M_s$): puts an upper bound on
  $g_{qV}^{33} \sin \theta_D \cos \theta_D$
[Eq.~(\ref{BsmixingVB})].

\item $\tau \to 3\mu$: puts an upper bound on $(g_{\ell V}^{33})^2
  \sin^3 \theta_L \cos \theta_L$ [Eq.~(\ref{tau3muamp})].

\item $\bsmumu$ ($C_9^{\mu\mu}({\rm NP})$): puts a lower bound on $(
  g_{qV}^{33} \sin \theta_D \cos \theta_D ) ( g_{\ell V}^{33} \sin^2
  \theta_L )$. (There is also an upper bound, but this is not relevant
  for the $VB$ model.)

\end{enumerate}
These three constraints overlap at basically a single point in the
parameter space. However, one still has the freedom to relabel this
point by adjusting the values of $g_{qV}^{33}$, $g_{\ell V}^{33}$,
$\theta_L$ and $\theta_D$. For example, in the previous section we had
$g_{qV}^{33} = g_{\ell V}^{33} = \sqrt{0.5}$, $(\theta_L, \theta_D) =
(0.333,0.006)$. However, two other possibilities are $g_{qV}^{33} =
\sqrt{0.5}/0.8,~ g_{\ell V}^{33} = 0.8\sqrt{0.5}$, $(\theta_L,
\theta_D) = (0.392,0.005)$ and $g_{qV}^{33} = \sqrt{0.28}/1.2,~ g_{\ell
  V}^{33} = 1.2 \sqrt{0.28}$, $(\theta_L, \theta_D) =
(0.360,0.009)$. But the key point is that, in both of these cases, the
predictions for other processes are little changed from those in
Sec.~\ref{Pred_processes}.

\item The $U_1$ model is viable only if $\l| h_{U_1}^{33}\r|^2 \ge
  0.5$. Values of the coupling$^2$ larger than 5 are allowed, see
  Fig.~\ref{Fig:U1coupvary}.

\begin{figure}[t]
\begin{center}
\includegraphics[viewport=0 0 360 395, width=16em]{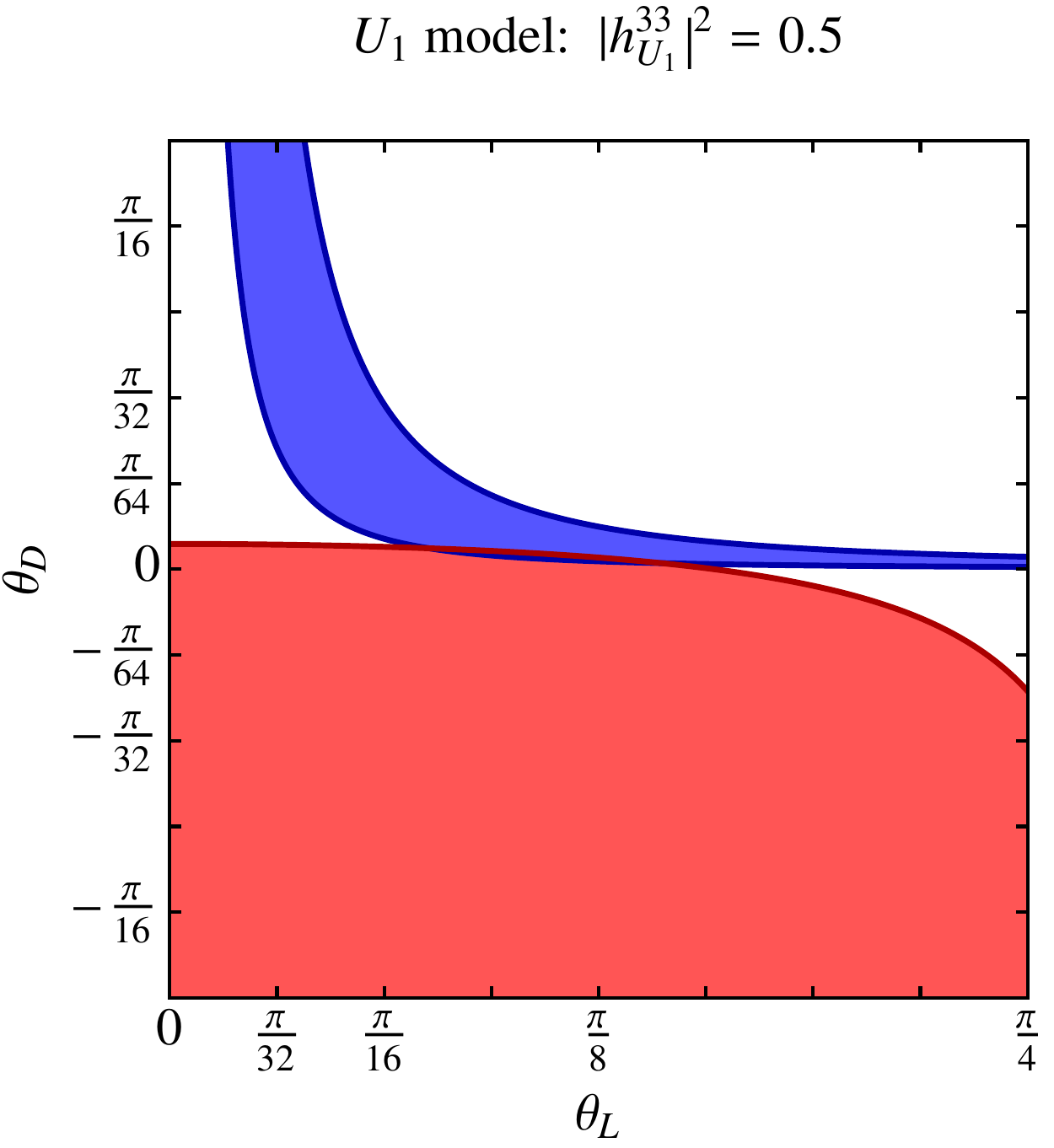}~~~~
\includegraphics[viewport=0 0 360 395, width=16em]{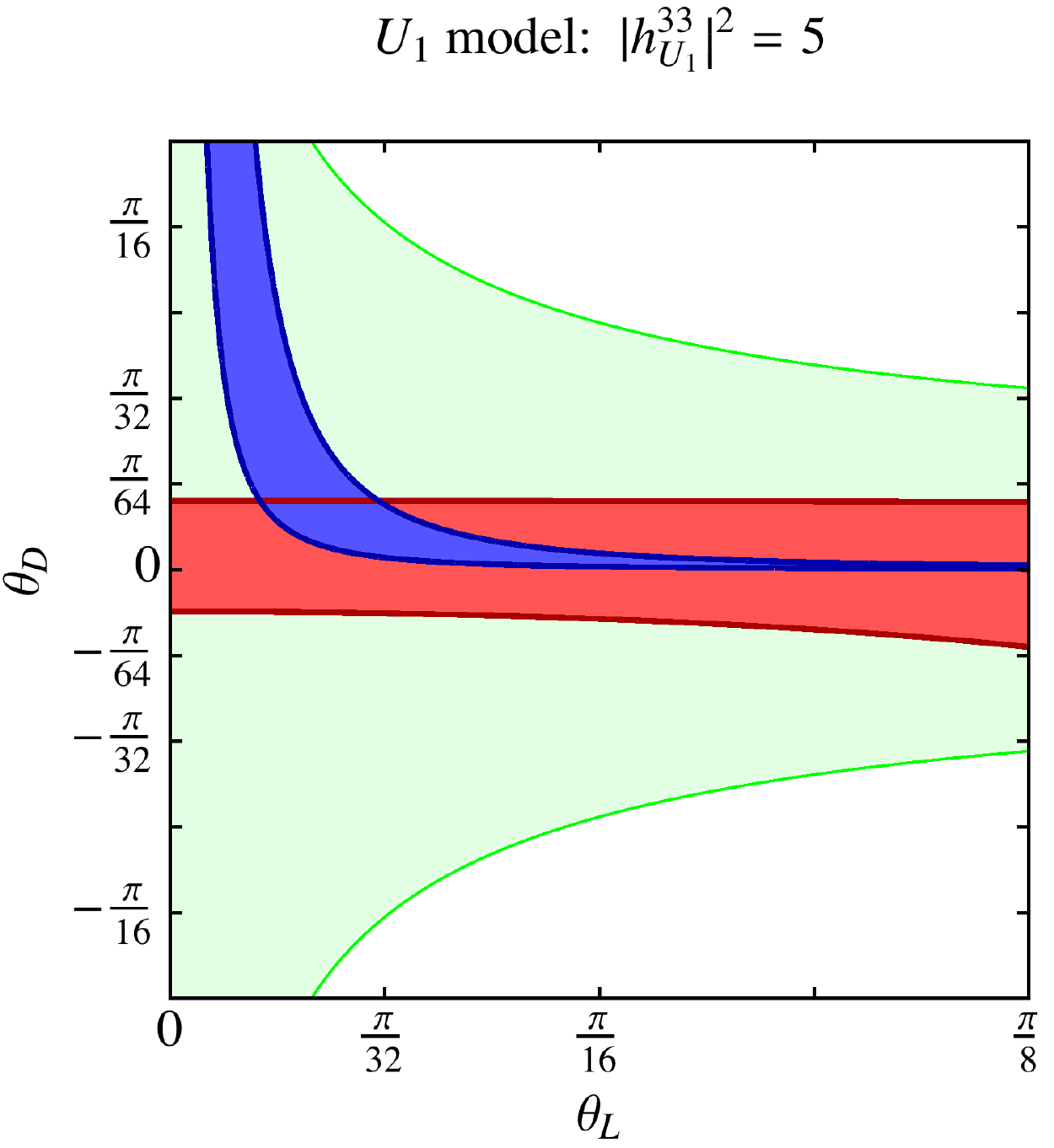}
\caption{Fig.~2 for the $U_1$ model, but with $\l| h_{U_1}^{33}\r|^2 =
  0.5$ (left) or $\l| h_{U_1}^{33}\r|^2 = 5$ (right). Left: the blue
  ($R_K$) and red ($R_{D^*}$) regions barely overlap, so this is the
  minimum value of the coupling$^2$ for which $U_1$ is viable. Right:
  the regions overlap, so $U_1$ is viable for $\l| h_{U_1}^{33}\r|^2 =
  5$ (as well as for larger values of the coupling$^2$).}
\label{Fig:U1coupvary}
\end{center}
\end{figure}

\end{itemize}

In fact, we do have some information about the value of the
coupling$^2$.  One can set limits on ${\rm coupling}^2/\Lambda^2_\NP$
from direct searches, assuming a certain mode of production for the
new mediator states. Following Ref.~\cite{direct_limit}, using the $ b
\bar{b} \to \tau \bar{\tau}$ process mediated by $s$- or $t$-channel
vector-boson or leptoquark exchange, one can get the following rough
upper bounds: $|g^{33}_{qV} g^{33}_{\ell V}|_{\rm max} / \Lambda^2_\NP
\sim 3$ TeV$^{-2}$ for the $VB$ model and $\l| h_{U_1}^{33}\r|^2_{\rm
  max} /\Lambda^2_\NP \sim 5$ TeV$^{-2}$ for the $U_1$ model. That is,
for $\Lambda_\NP = 1$ TeV, $g^{33}_{qV} g^{33}_{\ell V} \le 3$ and
$\l| h_{U_1}^{33}\r|^2 \le 5$\footnote{To be precise, the bound given
  in Ref.~\cite{direct_limit} should be applied as $g^{33}_{qV}
  g^{33}_{\ell V} \cos^2\theta_D \cos^2\theta_L \le 3$ and $\l|
  h_{U_1}^{33} \cos\theta_D \cos\theta_L \r|^2 \le 5$ (for
  $\Lambda_\NP = 1$ TeV).  The down-sector mixing, which reduces the
  rate of $b \bar b$ pair production, is negligible since $\theta_D
  \ll 1$ for the present case. As for the lepton mixing, it can at
  most reduce the decay rate into $\tau\bar\tau$ by 15\% (for
  $\theta_L \le \pi/8$).  Here we (conservatively) ignore this effect,
  resulting in a slightly more stringent constraint on ${\rm
    coupling}^2$, as shown in the main text.}.

In light of these results, we rederive the predictions of the
$U_1$ model for the various observables,
allowing $0.5 \le \l| h_{U_1}^{33}\r|^2 \le 5$. For comparison, we include
  the $VB$ predictions from Sec.~\ref{Pred_processes}.  We find
\begin{enumerate}

\item $R_{D^{(*)}}$:
\bea
VB &~~:~~& R_{D^{(*)}}^{\rm ratio} \simeq 1.04 ~, \nn\\
U_1 &~~:~~& 1.02 \le R_{D^{(*)}}^{\rm ratio} \le 1.29 ~.
\eea
For ${\rm coupling}^2
  = 1$, we found that, for both models, large deviations in
  $R_{D^{(*)}}$ from the SM are not favored, so it is not possible to
  distinguish the $VB$ and $U_1$ models using $R_{D^{(*)}}$. From the
  above numbers, we see that, when the coupling$^2$ is allowed to
  vary, this no longer holds. If it is found that $1.04 < R_{D^{(*)}}
  \le 1.29$, this will indicate $U_1$.

\item $R_K$:
\bea
VB &~~:~~& R_K \simeq 0.90 ~, \nn\\
U_1 &~~:~~& 0.51 \le R_K \le 0.90 ~.
\eea
The result is as
  before: if future measurements find $0.51 \le R_K < 0.90$, this
  would point clearly to $U_1$ (and exclude $VB$).


\item $B \to K^{(*)} \mu \tau$:
\bea
VB &~~:~~& \cB(B \to K^{(*)} \mu \tau) \simeq 4.0 \times 10^{-10} ~, \nn\\
U_1 &~~:~~& \cB(B \to K^{(*)} \mu \tau)|_{\rm max} = 1.6 \times 10^{-7} ~.
\eea
When the
  coupling$^2$ is allowed to vary, the value of $\cB(B \to K^{(*)} \mu
  \tau)|_{\rm max}$ predicted by the $U_1$ model is larger than in
  Sec.~\ref{Pred_processes}. Unfortunately, it is still below the
  reach of Belle II (which is $5 \times 10^{-7}$
  \cite{Belleprivcomm}).

\item $B \to K^{(*)} \tau^+ \tau^-$:
\bea
VB &~~:~~& \cB(B \to K^{(*)} \tau^+ \tau^-) \simeq 4.4 \times 10^{-8} ~, \nn\\
U_1 &~~:~~& \cB(B \to K^{(*)} \tau^+ \tau^-)|_{\rm max} = 1.1 \times 10^{-4} ~.
\eea
Here too, when the coupling$^2$ is allowed to vary, we find that the
value of $\cB(B \to K^{(*)} \tau^+ \tau^-)|_{\rm max}$
  for the $U_1$ model is increased over that in
  Sec.~\ref{Pred_processes}.  It may just be attainable at Belle II
(its reach is $\sim 2 \times 10^{-4}$ \cite{Belleprivcomm}). Thus, $B
\to K^{(*)} \tau^+ \tau^-$ could perhaps be used to distinguish the
two models.

\item $\bs \to \mu\tau$:
\bea
VB &~~:~~& \cB(\bs \to \mu\tau) \simeq 6.7 \times 10^{-9} ~, \nn\\
U_1 &~~:~~& \cB(\bs \to \mu\tau)|_{\rm max} = 2.8 \times 10^{-6} ~.
\eea
Once again, the value of $\cB(\bs \to \mu\tau)|_{\rm max}$ for the $U_1$ model is
  larger than that in Sec.~\ref{Pred_processes}. However, we cannot
evaluate whether this decay can be used to distinguish the two models
as we do not know the reach of LHCb or Belle II for $\bs \to \mu\tau$.

\item $\bs \to \tau^+ \tau^-$
\bea
VB &~~:~~& \cB(\bs \to \tau^+ \tau^-) \simeq 2.4 \times 10^{-7} ~, \nn\\
U_1 &~~:~~& \cB(\bs \to \tau^+ \tau^-)|_{\rm max} = 5.4 \times 10^{-4} ~.
\eea
The value of $\cB(\bs \to \tau^+ \tau^-)|_{\rm max}$ for the $U_1$ model is larger
  than before.  However, we cannot evaluate whether this decay can be
used to distinguish the two models as we do not know the reach of LHCb
or Belle II for $\bs \to \tau^+ \tau^-$.

\item $\Upsilon(3S) \to \mu \tau$:
\bea
VB && \cB(\Upsilon(3S) \to \mu \tau) \simeq 3.0 \times 10^{-9} ~, \nn\\[0.5em]
U_1 &~~:~~& \cB(\Upsilon(3S) \to \mu \tau)|_{\rm max} = 8.0 \times 10^{-7} ~.
\eea
Previously, we made a rough estimate that Belle II should be able to
measure $\cB(\Upsilon(3S) \to \mu \tau)$ down to $\sim 10^{-7}$. We
speculated that perhaps Belle II could do better than this (and noted
that a precise determination of the reach can only be obtained through
a detailed study). However, the above predicted values of
$\cB(\Upsilon(3S) \to \mu \tau)|_{\rm max}$ show that, even with our
rough estimate, the $U_1$ model can lead to rates for $\Upsilon(3S)
\to \mu \tau$ that are easily observable at Belle II. If this decay
were seen, it would exclude $VB$ and point to $U_1$. This demonstrates
the importance of this process for testing NP models in $B$ decays.

\end{enumerate}

\subsection{Combining Observables}

Above, we have seen that it is indeed possible to distinguish the $VB$
and $U_1$ models. $VB$ predicts that $\tau \to 3\mu$ is on the verge of being observed, while there are
several other observables that are signals of $U_1$. Should one of these
signals be seen, indicating the presence of a particular type of NP, it
would of course be very exciting. However, even more information about
the underlying NP model can be obtained by using the measurements of
other observables.

The $U_1$ model contains three unknown parameters: $\theta_L$, $\theta_D$
and $|h^{33}_{U_1}|^2/\Lambda^2_\NP$. Then, given the measurement of an
observable that indicates the presence of the $U_1$ model, one can use
two other observables to derive the values of all the parameters of the
model. To illustrate this, suppose that $R_K$ and $R_{D^{(*)}}$ are
measured very precisely, and $R_K = 0.781$ and $R^{\rm ratio}_{D^{(*)}}
= 1.077$ are found. If $\cB(\Upsilon(3S) \to \mu \tau) = 1.11 \times
10^{-8}$ is also measured, this points to the $U_1$ model. The theoretical
parameters must take the values $|h^{33}_{U_1}|^2/\Lambda^2_\NP = 2.43\,\text{TeV}^{-2}$,
$\theta_L = 0.039$, $\theta_D = 0.006$. $U_1$ then predicts $\cB(B \to
K^{(*)} \mu \tau) = 1.68 \times 10^{-8}$, $\cB(B \to K^{(*)} \tau^+ \tau
^-) = 5.57 \times 10^{-6}$, $\cB(\bs \to \mu\tau) = 2.80 \times 10^{-7}$,
and $\cB(\bs \to \tau^+ \tau^-) = 2.57 \times 10^{-5}$.

On the other hand, the $VB$ model is much more restrictive. It contains
four unknown parameters: $\theta_L$, $\theta_D$, $g^{33}_{qV}$ and $g^{33}
_{lV}$ (without loss of generality we can set $\Lambda_\NP = 1$ TeV).
Unlike the $U_1$ model, the $VB$ model receives severe constraints from
$\bs$-$\bsbar$ mixing and $\tau\to3\mu$. The constraint from $\bs$-$\bsbar$
mixing implies $g_{qV}^{33} \sin \theta_D \cos \theta_D < 4.2 \times 10^{-3
}$, while the constraint from $\tau\to3\mu$ implies $(g_{\ell V}^{33})^2
  \sin^3 \theta_L \cos \theta_L < 1.65 \times 10^{-2}$ at 90\% C.L. These
constraints leave no room for the $VB$ model to explain $R_K$ less than
0.90 at 90\% or higher C.L.

\section{Conclusions}

At present, there are several measurements of $B$ decays that exhibit
discrepancies with the predictions of the SM. These include $P'_5$
(from an angular analysis of $B \to K^* \mu^+\mu^-$), the differential
branching fraction of $B_s^0 \to \phi \mu^+ \mu^-$, $R_K \equiv {\cal
  B}(B^+ \to K^+ \mu^+ \mu^-)/{\cal B}(B^+ \to K^+ e^+ e^-)$, and
$R_{D^{(*)}} \equiv {\cal B}(\bar{B} \to D^{(*)} \tau^{-}
{\bar\nu}_\tau)/{\cal B}(\bar{B} \to D^{(*)} \ell^{-} {\bar\nu}_\ell)$
($\ell = e,\mu$). These suggest NP in $\bbtosb \mu^+ \mu^-$ (first
three signals) or ${\bar b} \to {\bar c} \tau^+ \nu_\tau$
($R_{D^{(*)}}$). Now, suppose that NP is present, and that it couples
preferentially to the left-handed third-generation particles in the
gauge basis. In Ref.~\cite{RKRD}, it was noted that, if this NP is
invariant under the full $SU(3)_C \times SU(2)_L \times U(1)_Y$ gauge
group, then, when one transforms to the mass basis, one generates the
operators $({\bar b}_L \gamma_\mu s_L) ({\bar \mu}_L \gamma^\mu \mu_L)$
(that contributes to $\bbtosb \mu^+ \mu^-$) {\it and} $({\bar b}_L
\gamma_\mu c_L) ({\bar \tau}_L \gamma^\mu \nu_{\tau L})$ (that
contributes to ${\bar b} \to {\bar c} \tau^+ \nu_\tau$). In other
words, the $R_K$ and $R_{D^{(*)}}$ puzzles can be simultaneously
explained.

This idea was explored in greater detail, using an effective field
theory approach, in Ref.~\cite{EffFT_3rdgen}. Here the starting point
is a model-independent effective Lagrangian consisting of two
four-fermion operators in the gauge basis, each with its own
coupling. It was assumed that the transformation from the gauge basis
to the mass basis leads to mixing only between the second and third
generations. As a consequence, for the down-type quarks, only one
unknown theoretical parameter is introduced: $\theta_D$. Similarly,
for the charged leptons, $\theta_L$ is the new parameter. In the mass
basis, the two operators contribute to a variety of $B$ decays, all
with two quarks and two leptons at the fermion level: $B \to K^*
\mu^+\mu^-$, $B_s^0 \to \phi \mu^+ \mu^-$, $R_K$, $R_{D^{(*)}}$, $B
\to K^{(*)} \nu {\bar\nu}$. The coefficients of the operators in the
mass basis are all functions of the coupling$^2$, $\theta_D$ and
$\theta_L$. For assumed values of the coupling$^2$, the experimental
measurements lead to an allowed region in $(\theta_L,\theta_D)$
space. This region was found to be nonzero, showing that a
simultaneous explanation of $R_K$ and $R_{D^{(*)}}$ is possible. There
are two UV completions that can give rise to the effective
Lagrangian. They are (i) $VB$: a vector boson that transforms as an
$SU(2)_L$ triplet, as in the SM, and (ii) $U_1$: an $SU(2)_L$-singlet
vector leptoquark.

The purpose of this paper is to explore ways of distinguishing the
$VB$ and $U_1$ models. There are two reasons to think that this might
be possible. First, the $VB$ model does not lead only to tree-level
operators with two quarks and two leptons. It also produces four-quark
and four-lepton operators. As such, it also contributes to processes
such as $\bs$-$\bsbar$ mixing and $\tau \to 3\mu$. These will lead to
additional constraints on $\theta_D$ and $\theta_L$, respectively.
Second, while $VB$ contributes to $B \to K^{(*)} \nu {\bar\nu}$, $U_1$
does not. The net effect is that the experimental constraints on the
$VB$ model are more stringent than those on the $U_1$ model. That is,
the allowed region in $(\theta_L,\theta_D)$ space is smaller for $VB$
than for $U_1$. This implies that the predictions for the rates of
other lepton-flavor-violating processes may be very different in the
two models, which will allow us to distinguish them.

With this in mind, our first step was to apply the relevant
experimental constraints to determine the allowed region in
$(\theta_L,\theta_D)$ space for each of the models. The constraints
from the measurements of $R_K$, $R_D$, $R_{D^*}$, and $\tau\to\mu\phi$
applied to both models.  For $VB$ there were additional constraints
from $B \to K^{(*)} \nu {\bar\nu}$, $\bs$-$\bsbar$ mixing, and $\tau
\to 3\mu$.

Our intention was to then use the allowed $(\theta_L,\theta_D)$
regions to compute the predictions of the two models for various
observables. However, the first step produced an unexpected result:
the constraints on the $VB$ model are so stringent that it is just
barely viable. To be specific, the boundaries of the allowed
$(\theta_L,\theta_D)$ regions corresponding to the $\Delta M_s$, $b
\to s\mu\mu$, and $\tau\to 3\mu$ constraints overlap at essentially a
single point. This is a very different result than that found in the
effective field theory analysis of Ref.~\cite{EffFT_3rdgen}.  This is
because all constraints have been included in the present model-dependent
analysis.  This illustrates that the results from the effective field
theory analysis must be used carefully: despite being
``model-independent,'' they do not necessarily apply to all models.

Things were very different for the $U_1$ model. We considered two
possibilities for the the coupling$^2$: $\l| h_{U_1}^{33}\r|^2 = 1$
and $0.5 \le \l| h_{U_1}^{33}\r|^2 \le 5$. In either case, the allowed
region in $(\theta_L,\theta_D)$ space is sizeable.

For both models, using the allowed $(\theta_L,\theta_D)$ regions, we
then computed the predictions for various observables. Note that, since the
$(\theta_L,\theta_D)$ ``region'' of the $VB$ model consists
essentially of a single point, the predictions for the observables are
very specific. On the other hand, the $U_1$ model gives ranges for its
predictions. The observables include $R_{D^{(*)}}$, $R_K$, $\tau \to
3\mu$ ($VB$ only), $B \to K^{(*)} \mu \tau$, $B \to K^{(*)} \tau^+
\tau^-$, $\bs \to \mu\tau$, $\bs \to \tau^+ \tau^-$ and $\Upsilon(3S)
\to \mu \tau$.  Note that the lepton-flavor-violating decay
$\Upsilon(3S) \to \mu \tau$ has been overlooked in previous
analyses. However, it is potentially an important process for testing
models proposed to explain the $B$-decay anomalies.

Given that their allowed $(\theta_L,\theta_D)$ regions are so
different, it is indeed possible to distinguish the $VB$ and $U_1$
models experimentally. $VB$ predicts $\cB(\tau^-\to\mu^-\mu^+\mu^-)
\simeq 2.1 \times 10^{-8}$, which is the present upper limit.  This is
measurable at Belle II and LHCb, so that $\tau \to 3\mu$ constitutes a
smoking-gun signal for the $VB$ model. There is no similar observable
for the $U_1$ model. However, there are a number of processes that can
potentially point to $U_1$. We present the results for $0.5 \le \l|
h_{U_1}^{33}\r|^2 \le 5$. For the decay $\Upsilon(3S) \to \mu \tau$,
we estimated that Belle II should be able to measure its branching
ratio down to $\sim 10^{-7}$. But the $U_1$ ($VB$) model predicts
$\cB(\Upsilon(3S) \to \mu \tau)|_{\rm max} = 8.0 \times 10^{-7}$ ($3.0
\times 10^{-9}$).  Thus, if this decay were observed, it would
indicate $U_1$ (and exclude $VB$). Another possibility is $R_K$. Its
present allowed $3\sigma$ range is $0.498 \le R_K \le 1.036$. The
$U_1$ ($VB$) model predicts $0.51 \le R_K \le 0.90$ ($R_K \simeq
0.90$).  The $U_1$ model can therefore accomodate smaller values of
$R_K$ than can the $VB$ model, so that, if future measurements find
$0.51 \le R_K < 0.90$ at higher than 90\% C.L.,
this would exclude $VB$ and favor $U_1$. Finally, for the other
decays $B \to K^{(*)} \mu \tau$, $B \to K^{(*)} \tau^+\tau^-$, $\bs
\to \mu\tau$, and $\bs \to \tau^+ \tau^-$, in all cases the $U_1$
model predicts larger branching ratios than does $VB$. However, whether
or not these decays can be used to distinguish the two models depends
on whether they can be observed at Belle II or LHCb.

\bigskip
\noindent
{\bf Notes Added}: (1) While this paper was being completed, the Belle
Collaboration released a new measurement of $R_{D^{*}}$
\cite{newBelleRD}. They find consistency with the SM at the level of
$0.6\sigma$. Now, if this result is combined with the previous results
of BaBar, Belle and LHCb, the discrepancy with the SM is reduced.
However, in any case, neither of the $VB$ and $U_1$ models presented
in this paper allows for large deviations in $R_{D^{(*)}}$ from the
SM.  Thus, this result is rather favored. (2) After this paper was
submitted to the arXiv, we were informed that LHCb has now set the
upper limit $\cB(\bs \to \tau^+ \tau^-) < 3.0 \times 10^{-3}$ (95\%
C.L.) \cite{BstautauLHCb}.

\bigskip
\noindent
{\bf Acknowledgments}: We thank S. Robertson and P. Urquijo for
information about the reach of Belle II, and E. Ben-Haim, T. Gershon,
F. Polci and J.  Serrano for information about the reach of LHCb. We
thank some people for bringing to our attention certain references:
G. Isidori (Refs.~\cite{IsidoriRK,U1LQ}), J.  Serrano
(Ref.~\cite{BstautauLHCb}).  DL thanks R. Chouiab and S. Robertson for
helpful discussions about the experimental measurement of $B \to K
\tau^+ \tau^-$. RW thanks A. Ishikawa for discussions about the
experimental measurement of $B_s \to \tau^+ \tau^-$ at Belle~II.
RW is grateful to Kenji Nishiwaki for giving us a critical question
so that we could discover an error in our fitting program.
This work was financially supported by NSERC of Canada (BB, JPG, DL), and
by the National Science Foundation (AD) under Grant No.\ NSF
PHY-1414345. BB acknowledges partial support from the
U.\ S.\ Department of Energy under contract DE-SC0007983. This work
was supported by IBS under the project code, IBS-R018-D1 (RW). AD
acknowledges the hospitality of the Department of Physics and
Astronomy, University of Hawaii, where part of the work was done.


\begin{thebibliography}{99}

\bibitem{BK*mumuLHCb1}
R.~Aaij {\it et al.} [LHCb Collaboration],
  ``Measurement of Form-Factor-Independent Observables in the Decay $B^{0} \to K^{*0} \mu^+ \mu^-$,''
  Phys.\ Rev.\ Lett.\  {\bf 111}, 191801 (2013)
  doi:10.1103/PhysRevLett.111.191801
  [arXiv:1308.1707 [hep-ex]].

\bibitem{BK*mumuLHCb2}
R.~Aaij {\it et al.} [LHCb Collaboration],
  ``Angular analysis of the $B^{0} \to K^{*0} \mu^{+} \mu^{-}$ decay using 3 fb$^{-1}$ of integrated luminosity,''
  JHEP {\bf 1602}, 104 (2016)
  doi:10.1007/JHEP02(2016)104
  [arXiv:1512.04442 [hep-ex]].

\bibitem{BK*mumuSM} U.~Egede, T.~Hurth, J.~Matias, M.~Ramon and
  W.~Reece, ``New observables in the decay mode ${\bar B}_d \to {\bar
    K}^{*0} l^+ l^-$,''
  JHEP {\bf 0811}, 032 (2008)
  doi:10.1088/1126-6708/2008/11/032 [arXiv:0807.2589 [hep-ph]].

\bibitem{BK*mumuBelle}
A.~Abdesselam {\it et al.} [Belle Collaboration],
  ``Angular analysis of $B^0 \to K^\ast(892)^0 \ell^+ \ell^-$,''
  arXiv:1604.04042 [hep-ex].

\bibitem{P'5} S.~Descotes-Genon, T.~Hurth, J.~Matias and J.~Virto,
  ``Optimizing the basis of $B \to K^* l l$ observables in the full kinematic range,''
  JHEP {\bf 1305}, 137 (2013)
  doi:10.1007/JHEP05(2013)137
  [arXiv:1303.5794 [hep-ph]].

\bibitem{BK*mumuhadunc1}
S.~Descotes-Genon, L.~Hofer, J.~Matias and J.~Virto,
  ``On the impact of power corrections in the prediction of $B \to K^*\mu^+\mu^-$ observables,''
  JHEP {\bf 1412}, 125 (2014)
  doi:10.1007/JHEP12(2014)125
  [arXiv:1407.8526 [hep-ph]].

\bibitem{BK*mumuhadunc2}
J.~Lyon and R.~Zwicky,
  ``Resonances gone topsy turvy - the charm of QCD or new physics in $b \to s \ell^+ \ell^-$?,''
  arXiv:1406.0566 [hep-ph].

\bibitem{BK*mumuhadunc3}
S.~J{\"a}ger and J.~Martin Camalich,
  ``Reassessing the discovery potential of the $B \to K^{*} \ell^+\ell^-$ decays in the large-recoil region: SM challenges and BSM opportunities,''
  Phys.\ Rev.\ D {\bf 93}, 014028 (2016)
  doi:10.1103/PhysRevD.93.014028
  [arXiv:1412.3183 [hep-ph]].

\bibitem{BK*mumuhaduncItalian}
M.~Ciuchini, M.~Fedele, E.~Franco, S.~Mishima, A.~Paul, L.~Silvestrini and M.~Valli,
  ``$B\to K^* \ell^+ \ell^-$ decays at large recoil in the Standard Model: a theoretical reappraisal,''
  arXiv:1512.07157 [hep-ph].

\bibitem{BK*mumulatestfit1}
S.~Descotes-Genon, L.~Hofer, J.~Matias and J.~Virto,
  ``Global analysis of $b\to s\ell\ell$ anomalies,''
  arXiv:1510.04239 [hep-ph].

\bibitem{BK*mumulatestfit2}
T.~Hurth, F.~Mahmoudi and S.~Neshatpour,
  ``On the anomalies in the latest LHCb data,''
  arXiv:1603.00865 [hep-ph].

\bibitem{BsphimumuLHCb1}
R.~Aaij {\it et al.} [LHCb Collaboration],
  ``Differential branching fraction and angular analysis of the decay $B_s^0\to\phi\mu^{+}\mu^{-}$,''
  JHEP {\bf 1307}, 084 (2013)
  doi:10.1007/JHEP07(2013)084
  [arXiv:1305.2168 [hep-ex]].

\bibitem{BsphimumuLHCb2}
R.~Aaij {\it et al.} [LHCb Collaboration],
  ``Angular analysis and differential branching fraction of the decay $B^0_s\to\phi\mu^+\mu^-$,''
  JHEP {\bf 1509}, 179 (2015)
  doi:10.1007/JHEP09(2015)179
  [arXiv:1506.08777 [hep-ex]].

\bibitem{latticeQCD1}
R.~R.~Horgan, Z.~Liu, S.~Meinel and M.~Wingate,
  ``Calculation of $B^0 \to K^{*0} \mu^+ \mu^-$ and $B_s^0 \to \phi \mu^+ \mu^-$ observables using form factors from lattice QCD,''
  Phys.\ Rev.\ Lett.\  {\bf 112}, 212003 (2014)
  doi:10.1103/PhysRevLett.112.212003
  [arXiv:1310.3887 [hep-ph]],

\bibitem{latticeQCD2}
  ``Rare $B$ decays using lattice QCD form factors,''
  PoS LATTICE {\bf 2014}, 372 (2015)
  [arXiv:1501.00367 [hep-lat]].

\bibitem{QCDsumrules}
A.~Bharucha, D.~M.~Straub and R.~Zwicky,
  ``$B\to V\ell^+\ell^-$ in the Standard Model from Light-Cone Sum Rules,''
  arXiv:1503.05534 [hep-ph].

\bibitem{RKexpt} R.~Aaij {\it et al.}  [LHCb Collaboration],
  ``Test of lepton universality using $B^{+}\rightarrow K^{+}\ell^{+}\ell^{-}$ decays,''
  Phys.\ Rev.\ Lett.\  {\bf 113}, 151601 (2014)
  [arXiv:1406.6482 [hep-ex]].

\bibitem{IsidoriRK}
M.~Bordone, G.~Isidori and A.~Pattori,
  ``On the Standard Model predictions for $R_K$ and $R_{K^*}$,''
  Eur.\ Phys.\ J.\ C {\bf 76}, no. 8, 440 (2016)
  doi:10.1140/epjc/s10052-016-4274-7
  [arXiv:1605.07633 [hep-ph]].

\bibitem{RD_BaBar}
  J.~P.~Lees {\it et al.} [BaBar Collaboration],
  ``Measurement of an Excess of $\bar{B} \to D^{(*)}\tau^- \bar{\nu}_\tau$ Decays and Implications for Charged Higgs Bosons,''
  Phys.\ Rev.\ D {\bf 88}, 072012 (2013)
  doi:10.1103/PhysRevD.88.072012
  [arXiv:1303.0571 [hep-ex]].

\bibitem{RD_Belle}
M.~Huschle {\it et al.} [Belle Collaboration],
  ``Measurement of the branching ratio of $\bar{B} \to D^{(\ast)} \tau^- \bar{\nu}_\tau$ relative to $\bar{B} \to D^{(\ast)} \ell^- \bar{\nu}_\ell$ decays with hadronic tagging at Belle,''
  Phys.\ Rev.\ D {\bf 92}, 072014 (2015)
  doi:10.1103/PhysRevD.92.072014
  [arXiv:1507.03233 [hep-ex]].

\bibitem{RD_LHCb}
R.~Aaij {\it et al.} [LHCb Collaboration],
  ``Measurement of the ratio of branching fractions $\mathcal{B}(\bar{B}^0 \to D^{*+}\tau^{-}\bar{\nu}_{\tau})/\mathcal{B}(\bar{B}^0 \to D^{*+}\mu^{-}\bar{\nu}_{\mu})$,''
  Phys.\ Rev.\ Lett.\  {\bf 115}, 111803 (2015)
  Addendum: [Phys.\ Rev.\ Lett.\  {\bf 115}, 159901 (2015)]
  doi:10.1103/PhysRevLett.115.159901, 10.1103/PhysRevLett.115.111803
  [arXiv:1506.08614 [hep-ex]].

\bibitem{Dumont:2016xpj}
B.~Dumont, K.~Nishiwaki and R.~Watanabe,
  ``LHC constraints and prospects for $S_1$ scalar leptoquark explaining the $\bar B \to D^{(*)} \tau \bar\nu$ anomaly,''
  arXiv:1603.05248 [hep-ph].

\bibitem{Tanaka:2012nw}
M.~Tanaka and R.~Watanabe,
  ``New physics in the weak interaction of $\bar B\to D^{(*)}\tau\bar\nu$,''
  Phys.\ Rev.\ D {\bf 87}, 034028 (2013)
  doi:10.1103/PhysRevD.87.034028
  [arXiv:1212.1878 [hep-ph]].

  \bibitem{Datta:2013kja}
  A.~Datta, M.~Duraisamy and D.~Ghosh,
  ``Explaining the $B \to K^\ast \mu^+ \mu^-$ data with scalar interactions,''
  Phys.\ Rev.\ D {\bf 89}, 071501 (2014)
  doi:10.1103/PhysRevD.89.071501
  [arXiv:1310.1937 [hep-ph]].

\bibitem{HS1}
G.~Hiller and M.~Schmaltz,
  ``$R_K$ and future $b \to s \ell \ell$ physics beyond the standard model opportunities,''
  Phys.\ Rev.\ D {\bf 90}, 054014 (2014)
  doi:10.1103/PhysRevD.90.054014
  [arXiv:1408.1627 [hep-ph]].

\bibitem{globalfits1}
D.~Ghosh, M.~Nardecchia and S.~A.~Renner,
``Hint of Lepton Flavour Non-Universality in $B$ Meson Decays,''
  JHEP {\bf 1412}, 131 (2014)
  doi:10.1007/JHEP12(2014)131
  [arXiv:1408.4097 [hep-ph]].

\bibitem{globalfits2}
T.~Hurth, F.~Mahmoudi and S.~Neshatpour,
``Global fits to $b \to s\ell\ell$ data and signs for lepton non-universality,''
  JHEP {\bf 1412}, 053 (2014)
  doi:10.1007/JHEP12(2014)053
  [arXiv:1410.4545 [hep-ph]].

\bibitem{globalfits3}
W.~Altmannshofer and D.~M.~Straub,
  ``New physics in $b\rightarrow s$ transitions after LHC run 1,''
  Eur.\ Phys.\ J.\ C {\bf 75}, no. 8, 382 (2015)
  doi:10.1140/epjc/s10052-015-3602-7
  [arXiv:1411.3161 [hep-ph]].

\bibitem{GGL} S.~L.~Glashow, D.~Guadagnoli and K.~Lane,
  ``Lepton Flavor Violation in $B$ Decays?,''
  Phys.\ Rev.\ Lett.\  {\bf 114}, 091801 (2015)
  doi:10.1103/PhysRevLett.114.091801
  [arXiv:1411.0565 [hep-ph]].

\bibitem{RKRD}
  B.~Bhattacharya, A.~Datta, D.~London and S.~Shivashankara,
  ``Simultaneous Explanation of the $R_K$ and $R(D^{(*)})$ Puzzles,''
  Phys.\ Lett.\ B {\bf 742}, 370 (2015)
  doi:10.1016/j.physletb.2015.02.011
  [arXiv:1412.7164 [hep-ph]].

\bibitem{RD1}
  A.~Datta, M.~Duraisamy and D.~Ghosh,
  ``Diagnosing New Physics in $b \to c \, \tau \, \nu_\tau$ decays in the light of the recent BaBar result,''
  Phys.\ Rev.\ D {\bf 86}, 034027 (2012)
  doi:10.1103/PhysRevD.86.034027
  [arXiv:1206.3760 [hep-ph]].

  \bibitem{RD2}
  A.~Celis, M.~Jung, X.~Q.~Li and A.~Pich,
  ``Sensitivity to charged scalars in $\boldsymbol{B\to D^{(*)}\tau\nu_\tau}$ and $\boldsymbol{B\to\tau\nu_\tau}$ decays,''
  JHEP {\bf 1301}, 054 (2013)
  doi:10.1007/JHEP01(2013)054
  [arXiv:1210.8443 [hep-ph]].

  \bibitem{RD3}
  A.~Crivellin, A.~Kokulu and C.~Greub,
  ``Flavor-phenomenology of two-Higgs-doublet models with generic Yukawa structure,''
  Phys.\ Rev.\ D {\bf 87}, 094031 (2013)
  doi:10.1103/PhysRevD.87.094031
  [arXiv:1303.5877 [hep-ph]].

  \bibitem{RD4}
  I.~Dor{\v s}ner, S.~Fajfer, N.~Ko{\v s}nik and I.~Ni{\v s}and{\v z}i\'c,
  ``Minimally flavored colored scalar in $\bar B \to D^{(*)} \tau \bar \nu$ and the mass matrices constraints,''
  JHEP {\bf 1311}, 084 (2013)
  doi:10.1007/JHEP11(2013)084
  [arXiv:1306.6493 [hep-ph]].

  \bibitem{RD5}
  M.~Freytsis, Z.~Ligeti and J.~T.~Ruderman,
  ``Flavor models for $\bar{B} \to D^{(*)} \tau \bar{\nu}$,''
  Phys.\ Rev.\ D {\bf 92}, 054018 (2015)
  doi:10.1103/PhysRevD.92.054018
  [arXiv:1506.08896 [hep-ph]].

  \bibitem{RD6}
  N.~G.~Deshpande and X.~G.~He,
  ``Consequences of R-Parity violating interactions for anomalies in $\bar B\to D^{(*)} \tau \bar \nu$ and $b\to s \mu^+\mu^-$,''
  arXiv:1608.04817 [hep-ph].

\bibitem{RD7}
  M.~A.~Ivanov, J.~G.~K\"orner and C.~T.~Tran,
  ``Analyzing new physics in the decays $\bar{B}^0 \to D^{(\ast)}\tau^-\bar\nu_{\tau}$
  with form factors obtained from the covariant quark model,''
  arXiv:1607.02932 [hep-ph].

  \bibitem{distributionD1}
  M.~Duraisamy and A.~Datta,
  ``The Full $B \to D^{*} \tau^{-} \bar{\nu_\tau}$ Angular Distribution and CP violating Triple Products,''
  JHEP {\bf 1309}, 059 (2013)
  doi:10.1007/JHEP09(2013)059
  [arXiv:1302.7031 [hep-ph]].

   \bibitem{distributionD2}
  M.~Duraisamy, P.~Sharma and A.~Datta,
  ``Azimuthal $B \to D^{*} \tau^{-} \bar{\nu_\tau}$ angular distribution with tensor operators,''
  Phys.\ Rev.\ D {\bf 90}, 074013 (2014)
  doi:10.1103/PhysRevD.90.074013
  [arXiv:1405.3719 [hep-ph]].

  \bibitem{distributionRW}
  Y.~Sakaki, M.~Tanaka, A.~Tayduganov, and R.~Watanabe,
  ``Probing New Physics with $q^2$ distributions in $\bar{B} \to D^{(*)} \tau \bar\nu$,''
  Phys.\ Rev.\ D {\bf 91}, 114028 (2015)
 doi:10.1103/PhysRevD.91.114028
  [arXiv:1412.3761 [hep-ph]].

\bibitem{anomalies1}
  D.~Das, C.~Hati, G.~Kumar and N.~Mahajan,
  ``Towards a unified explanation of $R_{D^{(\ast)}}$, $R_{K}$ and $(g-2)_{\mu}$ anomalies in a left-right model with leptoquarks,''
  Phys.\ Rev.\ D {\bf 94}, 055034 (2016)
  doi:10.1103/PhysRevD.94.055034
  [arXiv:1605.06313 [hep-ph]].

\bibitem{anomalies2}
  C.~J.~Lee and J.~Tandean,
  ``Minimal lepton flavor violation implications of the $b \to s$ anomalies,''
  JHEP {\bf 1508}, 123 (2015)
  doi:10.1007/JHEP08(2015)123
  [arXiv:1505.04692 [hep-ph]].

\bibitem{anomalies3}
  D.~Becirevic, S.~Fajfer and N.~Ko{\v s}nik,
  ``Lepton flavor nonuniversality in $b \to s l^+ l^-$ processes,''
  Phys.\ Rev.\ D {\bf 92}, 014016 (2015)
  doi:10.1103/PhysRevD.92.014016
  [arXiv:1503.09024 [hep-ph]].

\bibitem{EffFT_3rdgen}
L.~Calibbi, A.~Crivellin and T.~Ota,
  ``Effective Field Theory Approach to $b \to s \ell \ell^{(\prime)}$, $B \to K^{(*)} \nu {\bar\nu}$ and $B \to D^{(*)} \tau \nu$ with Third Generation Couplings,''
  Phys.\ Rev.\ Lett.\  {\bf 115}, 181801 (2015)
  doi:10.1103/PhysRevLett.115.181801
  [arXiv:1506.02661 [hep-ph]].

\bibitem{BKnunubarBaBar}
J.~P.~Lees {\it et al.} [BaBar Collaboration],
  ``Search for $B \to K^{(*)} \nu \overline \nu$ and invisible quarkonium decays,''
  Phys.\ Rev.\ D {\bf 87}, 112005 (2013)
  doi:10.1103/PhysRevD.87.112005
  [arXiv:1303.7465 [hep-ex]].

\bibitem{BKnunubarBelle}
O.~Lutz {\it et al.} [Belle Collaboration],
  ``Search for $B \to h^{(*)} \nu \bar{\nu}$ with the full Belle $\Upsilon(4S)$ data sample,''
  Phys.\ Rev.\ D {\bf 87}, 111103 (2013)
  doi:10.1103/PhysRevD.87.111103
  [arXiv:1303.3719 [hep-ex]].

\bibitem{AGC}
R.~Alonso, B.~Grinstein and J.~M.~Camalich,
  ``Lepton universality violation and lepton flavor conservation in $B$-meson decays,''
  JHEP {\bf 1510}, 184 (2015)
  doi:10.1007/JHEP10(2015)184
  [arXiv:1505.05164 [hep-ph]].

 \bibitem{Crivellin:2015lwa}
  A.~Crivellin, G.~D'Ambrosio and J.~Heeck,
  ``Addressing the LHC flavor anomalies with horizontal gauge symmetries,''
  Phys.\ Rev.\ D {\bf 91}, 075006 (2015)
  doi:10.1103/PhysRevD.91.075006
  [arXiv:1503.03477 [hep-ph]].

\bibitem{Isidori} A.~Greljo, G.~Isidori and D.~Marzocca,
  ``On the breaking of Lepton Flavor Universality in B decays,''
  JHEP {\bf 1507}, 142 (2015)
  doi:10.1007/JHEP07(2015)142
  [arXiv:1506.01705 [hep-ph]].

  \bibitem{dark}
  D.~Aristizabal Sierra, F.~Staub and A.~Vicente,
  ``Shedding light on the $b\to s$ anomalies with a dark sector,''
  Phys.\ Rev.\ D {\bf 92}, 015001 (2015)
  doi:10.1103/PhysRevD.92.015001
  [arXiv:1503.06077 [hep-ph]].

\bibitem{Chiang}
C.~W.~Chiang, X.~G.~He and G.~Valencia,
``$Z'$ model for $b \to s \ell {\bar\ell}$ flavor anomalies,''
  Phys.\ Rev.\ D {\bf 93}, 074003 (2016)
  doi:10.1103/PhysRevD.93.074003
  [arXiv:1601.07328 [hep-ph]].

\bibitem{Virto}
S.~M.~Boucenna, A.~Celis, J.~Fuentes-Martin, A.~Vicente and J.~Virto,
  ``Non-abelian gauge extensions for B-decay anomalies,''
  arXiv:1604.03088 [hep-ph].

\bibitem{RDLQs}
Y.~Sakaki, M.~Tanaka, A.~Tayduganov and R.~Watanabe,
  ``Testing leptoquark models in $\bar B \to D^{(*)} \tau \bar\nu$,''
  Phys.\ Rev.\ D {\bf 88}, 094012 (2013)
  doi:10.1103/PhysRevD.88.094012
  [arXiv:1309.0301 [hep-ph]].

\bibitem{S3LQ1}
B.~Gripaios, M.~Nardecchia and S.~A.~Renner,
  ``Composite leptoquarks and anomalies in $B$-meson decays,''
  JHEP {\bf 1505}, 006 (2015)
  doi:10.1007/JHEP05(2015)006
  [arXiv:1412.1791 [hep-ph]].

\bibitem{S3LQ2}
I.~de Medeiros Varzielas and G.~Hiller,
  ``Clues for flavor from rare lepton and quark decays,''
  JHEP {\bf 1506}, 072 (2015)
  doi:10.1007/JHEP06(2015)072
  [arXiv:1503.01084 [hep-ph]].

\bibitem{U1LQ}
R.~Barbieri, G.~Isidori, A.~Pattori and F.~Senia,
  ``Anomalies in $B$-decays and $U(2)$ flavour symmetry,''
  Eur.\ Phys.\ J.\ C {\bf 76}, no. 2, 67 (2016)
  doi:10.1140/epjc/s10052-016-3905-3
  [arXiv:1512.01560 [hep-ph]].

\bibitem{U3LQ}
S.~Fajfer and N.~Ko{\v s}nik,
 ``Vector leptoquark resolution of $R_K$ and $R_{D^{(*)}}$ puzzles,''
 Phys.\ Lett.\ B {\bf 755}, 270 (2016)
 doi:10.1016/j.physletb.2016.02.018
 [arXiv:1511.06024 [hep-ph]].

\bibitem{Sahoo:2016pet}
  S.~Sahoo, R.~Mohanta and A.~K.~Giri,
  ``Explaining $R_{K}$ and $R_{D^{(*)}}$ anomalies with vector leptoquark,''
  arXiv:1609.04367 [hep-ph].

\bibitem{Buras:2014fpa}
A.~J.~Buras, J.~Girrbach-Noe, C.~Niehoff and D.~M.~Straub,
  ``$ B\to {K}^{\left(\ast \right)}\nu \overline{\nu} $ decays in the Standard Model and beyond,''
  JHEP {\bf 1502}, 184 (2015)
  doi:10.1007/JHEP02(2015)184
  [arXiv:1409.4557 [hep-ph]].

\bibitem{Barate:2000rc}
R.~Barate {\it et al.} [ALEPH Collaboration],
  ``Measurements of $BR (b \to \tau^- {\bar\nu}_\tau X)$ and $BR ((b \to \tau^- {\bar\nu}_\tau D^{* \pm} X$)
  and upper limits on $BR (B^- \to \tau^- {\bar\nu}_\tau )$ and $BR (b \to s {\bar\nu})$,''
  Eur.\ Phys.\ J.\ C {\bf 19}, 213 (2001)
  doi:10.1007/s100520100612
  [hep-ex/0010022].

\bibitem{pdg}
  K.~A.~Olive {\it et al.} [Particle Data Group Collaboration],
  ``Review of Particle Physics,''
  Chin.\ Phys.\ C {\bf 38}, 090001 (2014).
  doi:10.1088/1674-1137/38/9/090001

  \bibitem{fetap}
  A.~Datta, X.~G.~He and S.~Pakvasa,
  ``Quasiinclusive and exclusive decays of B to eta-prime,''
  Phys.\ Lett.\ B {\bf 419}, 369 (1998)
  doi:10.1016/S0370-2693(97)01449-4
  [hep-ph/9707259].

\bibitem{etamix}
  B.~Bhattacharya and J.~L.~Rosner,
  ``Effect of $\eta$-$\eta'$ mixing on $D \rightarrow PV$ decays,''
  Phys.\ Rev.\ D {\bf 82}, 037502 (2010)
  doi:10.1103/PhysRevD.82.037502
  [arXiv:1005.2159 [hep-ph]].

\bibitem{taumuphiexp}
Y.~Miyazaki {\it et al.} [Belle Collaboration],
  ``Search for Lepton-Flavor-Violating tau Decays into a Lepton and a Vector Meson,''
  Phys.\ Lett.\ B {\bf 699}, 251 (2011)
  doi:10.1016/j.physletb.2011.04.011
  [arXiv:1101.0755 [hep-ex]].

\bibitem{Buchalla:1995vs}
G.~Buchalla, A.~J.~Buras and M.~E.~Lautenbacher,
  ``Weak decays beyond leading logarithms,''
  Rev.\ Mod.\ Phys.\  {\bf 68}, 1125 (1996)
  doi:10.1103/RevModPhys.68.1125
  [hep-ph/9512380].

\bibitem{Aoki:2013ldr}
S.~Aoki {\it et al.},
  ``Review of lattice results concerning low-energy particle physics,''
  Eur.\ Phys.\ J.\ C {\bf 74}, 2890 (2014)
  doi:10.1140/epjc/s10052-014-2890-7
  [arXiv:1310.8555 [hep-lat]].

\bibitem{Aoki:2016frl}
S.~Aoki {\it et al.},
  ``Review of lattice results concerning low-energy particle physics,''
  [arXiv:1607.00299 [hep-lat]].

\bibitem{Charles:2015gya}
J.~Charles {\it et al.},
  ``Current status of the Standard Model CKM fit and constraints on $\Delta F=2$ New Physics,''
  Phys.\ Rev.\ D {\bf 91}, 073007 (2015)
  doi:10.1103/PhysRevD.91.073007
  [arXiv:1501.05013 [hep-ph]].

\bibitem{Chetyrkin:2000yt}
K.~G.~Chetyrkin, J.~H.~Kuhn and M.~Steinhauser,
  ``RunDec: A Mathematica package for running and decoupling of the strong coupling and quark masses,''
  Comput.\ Phys.\ Commun.\  {\bf 133}, 43 (2000)
  doi:10.1016/S0010-4655(00)00155-7
  [hep-ph/0004189].

\bibitem{HFAG}
Y.~Amhis {\it et al.} [Heavy Flavor Averaging Group (HFAG) Collaboration],
  ``Averages of $b$-hadron, $c$-hadron, and $\tau$-lepton properties as of summer 2014,''
  arXiv:1412.7515 [hep-ex].

\bibitem{tau23muexp}
  K.~Hayasaka {\it et al.},
  ``Search for Lepton Flavor Violating Tau Decays into Three Leptons with 719 Million Produced Tau+Tau- Pairs,''
  Phys.\ Lett.\ B {\bf 687}, 139 (2010)
  doi:10.1016/j.physletb.2010.03.037
  [arXiv:1001.3221 [hep-ex]].

\bibitem{ParadisiEFT}
  F.~Feruglio, P.~Paradisi and A.~Pattori,
  ``Revisiting Lepton Flavour Universality in B Decays,''
  arXiv:1606.00524 [hep-ph].

\bibitem{flavio}
David Straub, \textit{flavio v0.11, 2016.}
  \href{http://dx.doi.org/10.5281/zenodo.59840}{http://dx.doi.org/10.5281/zenodo.59840}

\bibitem{Shintalk}
  Talk by Jing-Ge Shiu (National Taiwan University) on behalf of the
  Belle II collaboration, {\it Beauty 2016}, Marseille, France, \href{https://indico.cern.ch/event/352928/contributions/1757317/attachments/1267910/1877886/2016BEAUTY\_physjg.pdf}
  {https://indico.cern.ch/event/352928/contributions/1757317}.

\bibitem{LHCbLetterofIntent}
  LHCb Collaboration, ``Letter of Intent for the LHCb Upgrade,'' CERN-LHCC-2011-001, 2011.

\bibitem{BKmutauexp}
B.~Aubert {\it et al.} [BaBar Collaboration],
  ``Search for the decay $B^{+} \to K^{+} \tau^\mp \mu^\pm$,''
  Phys.\ Rev.\ Lett.\  {\bf 99}, 201801 (2007)
  doi:10.1103/PhysRevLett.99.201801
  [arXiv:0708.1303 [hep-ex]].

\bibitem{Belleprivcomm} Steven Robertson and Phillip Urquijo, private
  communication.

\bibitem{BaBarBKtautau}
[BaBar Collaboration],
  ``Search for $B^{+}\rightarrow K^{+} \tau^{+}\tau^{-}$ at the BaBar experiment,''
  arXiv:1605.09637 [hep-ex].

\bibitem{LHCbprivcomm} Tim Gershon, Francesco Polci and Justine
  Serrano, private communication.

  \bibitem{Datta:1998yq}
  A.~Datta, P.~J.~O'Donnell, S.~Pakvasa and X.~Zhang,
  ``Flavor changing processes in quarkonium decays,''
  Phys.\ Rev.\ D {\bf 60}, 014011 (1999)
  doi:10.1103/PhysRevD.60.014011
  [hep-ph/9812325].

\bibitem{BaBarUps}
  J.~P.~Lees {\it et al.} [BaBar Collaboration],
  ``Search for Charged Lepton Flavor Violation in Narrow Upsilon Decays,''
  Phys.\ Rev.\ Lett.\  {\bf 104}, 151802 (2010)
  doi:10.1103/PhysRevLett.104.151802
  [arXiv:1001.1883 [hep-ex]].

\bibitem{CLEOUps}
  W.~Love {\it et al.} [CLEO Collaboration],
  ``Search for Lepton Flavor Violation in Upsilon Decays,''
  Phys.\ Rev.\ Lett.\  {\bf 101}, 201601 (2008)
  doi:10.1103/PhysRevLett.101.201601
  [arXiv:0807.2695 [hep-ex]].

\bibitem{direct_limit}
  D.~A.~Faroughy, A.~Greljo and J.~F.~Kamenik,
  ``Confronting lepton flavor universality violation in B decays with high-$p_T$ tau lepton searches at LHC,''
  arXiv:1609.07138 [hep-ph].

\bibitem{newBelleRD} A.~Abdesselam {\it et al.},
  ``Measurement of the $\tau$ lepton polarization in the decay ${\bar B} \rightarrow D^* \tau^- {\bar \nu_{\tau}}$,''
  arXiv:1608.06391 [hep-ex].

\bibitem{BstautauLHCb} Talk by K. De Bruyn (LHCb Collaboration) at
  TAU2016, LHCb-CONF-2016-011. \href{http://indico.ihep.ac.cn/event/5221/session/12/contribution/69/material/slides/0.pdf}
  {http://indico.ihep.ac.cn/event/5221/session/12/contribution/69/material/slides/0.pdf}

\end{thebibliography}
\end{document}